# Measuring Global Multi-Scale Place Connectivity using Geotagged Social Media Data

Zhenlong Li[1*], Xiao Huang[2], Xinyue Ye[3], Yuqin Jiang[1], Yago Martin[4], Huan Ning[1],

Michael E. Hodgson[1], and Xiaoming Li[5]

[1] *Geoinformation and Big Data Research Laboratory, Department of Geography, University of South Carolina, SC, USA*

[2] *Department of Geosciences, University of Arkansas, AR, USA*

[3] *Department of Landscape Architecture & Urban Planning, Texas A&M University, TX, USA*

[4] *School of Public Administration, University of Central Florida, FL, USA*

[5] *Department of Health Promotion, Education, and Behavior, University of South Carolina, SC, USA*

[*] zhenlong@sc.edu

# Abstract

Shaped by human movement, place connectivity is quantified by the strength of spatial interactions among locations. For decades, spatial scientists have researched place connectivity, applications, and metrics. The growing popularity of social media provides a new data stream where spatial social interaction measures are largely devoid of privacy issues, easily assessable, and harmonized. In this study, we introduced a global multi-scale place connectivity index (PCI) based on spatial interactions among places revealed by geotagged tweets as a spatiotemporal-continuous and easy-to-implement measurement. The multi-scale PCI, demonstrated at the US county level, exhibits a strong positive association with SafeGraph population movement records (10% penetration in the US population) and Facebook's social connectedness index (SCI), a popular connectivity index based on social networks. We found that PCI has a strong boundary effect and that it generally follows the distance decay, although this force is weaker in more urbanized counties with a denser population. Our investigation further suggests that PCI has great potential in addressing real-world problems that require place connectivity knowledge, exemplified with two applications: 1) modeling the spatial spread of COVID-19 during the early stage of the pandemic and 2) modeling hurricane evacuation destination choice. The methodological and contextual knowledge of PCI, together with the launched visualization platform and open-sourced PCI datasets at various geographic levels, are expected to support research fields requiring knowledge in human spatial interactions.

# Introduction

Since the proposal of "social physics" in 1948 by John Stewart, an astrophysicist who first attempted to reveal spatial interaction based on the concept of the Newtonian gravitational framework (Stewart, 1948), research on modeling, documenting, and understanding human spatial interaction has been a research hotspot in geography and related fields. From a geographic perspective, human movements form the spatial interactions among places, featured by both social (population, land use, culture, etc.) and physical characteristics (climate, geology, landscape, etc.) (Massey, 1994). Relationships among places are shaped by constant human movement, and the intensity of such movement further quantifies the connectivity strength among places. Thus, understanding connectivity between two places provides fundamental knowledge regarding their interactive gravity, benefiting various applications such as infectious disease modeling, transportation planning, tourism management, evacuation modeling, and other fields requiring knowledge in human spatial interactions.

However, measuring such interactions at various spatiotemporal scales is a challenging task. Early efforts (widely adopted until now) to examine spatial interactions involved survey utilization. Researchers used questionnaires to understand spatial interactions, aiming to gauge both long-term spatial movement, such as migration patterns (Barcus and Brunn, 2010; Boyle, 2014; Salt, 1987;), and short-term spatial displacement, such as evacuation and traveling (Buliung and Kanaroglou, 2006; Calantone et al., 1989; Durage et al., 2014; Pham et al., 2020; Santos et al., 2011; Siebeneck and Cova, 2012). The well-documented spatial interactions from these surveys contribute to our understanding of how people move across space and how places are connected; however, such an approach suffers from limitations of small sample sizes (Martín et al., 2020a), limited temporal resolution (Pereira et al., 2013), and resource demands (Martín et al., 2020b).

The limitations of survey-based approaches largely preclude spatiotemporal-continuous observations in spatial interactions, therefore inducing discrete place connectivity measurements. However, place connectivity should not be considered as a fixed spatiotemporal property of places. Instead, connectivity is ever-changing and evolving rapidly in modern society (Batten, 2001; Gao, 2015; Macdonald and Grieco, 2007). As argued by many, technological advances in the past decades have greatly facilitated connectivity by weakening geographic limits (Tranos and Nijkamp, 2013). To capture the temporal nature of spatial interactions, researchers have emphasized the importance of transportation data that detail people's moving patterns. Place connectivity has been measured using various transportation means that include airline flows (Derudder and Witlox, 2008; Xu and Harriss, 2008), highway traffic (Zhang et al., 2020), railway flows (Yang et al., 2019; Lin et al., 2019), and intercity bus networks (Ghafoor et al., 2014). The rich traffic information and the derived spatial networks greatly facilitate our understanding of how places are connected via these transportation modes. However, transportation-based approaches pose new challenges. First, such data are generally difficult to obtain, as they are often confidential or collected by private companies. Second, the data themselves are mode-specific, lacking the holistic views of the overall human spatial interactions and place connectivity, which are often needed in fields such as infectious disease modeling. A notable effort to tackle the latter issue is by Lin et al. (2019), who constructed a combined inter-city connectivity measurement based on multiple data sources for nine cities in China and demonstrated its advantage over the index derived from a single data source. This study offers valuable insights in understanding how cities are connected using a holistic approach. Due to data availability issues, however, it is challenging to construct such a combined index that are spatiotemporal-continuous for a large area (e.g., a country or the entire world) at various geographic settings and scales (e.g., urban, suburb, or rural; county, state/province, or country).

The emerging concepts of "Web 2.0" (O'Reilly, 2007) and "Citizen as Sensors" (Goodchild, 2007), largely benefiting from the advent of geo-positioning technologies, offer a new avenue to actively and passively gather and collect the digital traces left by electronic device holders (Liu et al., 2015; Li et al., 2019). For example, passive trace collection involves data obtained from mobile phone data (Amini et al., 2014; Gonzalez et al., 2008), smart cards (Agard et al., 2006; Ma et al., 2013), or wireless networks (Perttunen et al., 2014). The spatial interactions documented from these passively collected traces tend to have high representativeness, given their high data penetration ratios. However, privacy and confidentiality concerns have been raised for such approaches, as individuals do not intend to actively share their locational information and are unaware of the usage of the generated positions (de Montjoye et al., 2018; Kontokosta and Johnson, 2017).

An approach less encumbered with privacy issues is based on spatial information from social media, a digital platform aiming to facilitate information sharing that has been popularized in recent years. Owing to their active sharing characteristics, social media data are less abundant compared to passively collected GPS positions from mobile devices but are less intrusive (Huang et al., 2020a; Jiang et al., 2019a), more accessible (Hu et al., 2020), and more harmonized (Huang et al., 2020b). The huge volume of user-generated content covering extensive areas facilitates the timely need for summarizing human spatial interactions. Twitter, for example, has quickly become the largest social media data source for geospatial research and has been widely used in human mobility studies (Fiorio et al., 2017; Jurdak et al., 2015; Li et al., 2020a; Soliman et al., 2017), given its free application programming interface (API) that allows unrestricted access to about 1% of the total tweets (Hawelka et al., 2014). We believe that the enormous sensing network constituted by millions of Twitter users worldwide provides unprecedented data to measure place connectivity at various spatiotemporal scales.

As an essential component in human interaction, social connections that involve online searching, friendships, account following, news mentioning, and information reposting can also contribute to place connectivity measurement. For example, the co-occurrences of toponyms on massive web documents, news articles, or social media were extracted to measure city relatedness and connectivity (Liu et al., 2014; Hu et al., 2017; Ye et al., 2020). A recent effort from Facebook explores connectivity measurement among places (called Social Connectedness Index, SCI) utilizing the social networks constructed from massive friendship links on Facebook (Bailey et al., 2018). However, whether or how the place connectivity measured by social connections differs from the one measured by physical connections is worth further investigation.

In view of the existing studies, gaps still exist in 1) the effort to construct a global place connectivity measurement that is harmonized, multi-scale, spatiotemporal-continuous based on the physical movement of social media users, 2) examining the utility of the derived place connectivity from a very large area and/or longer time period in solving some real-world problems, and 3) applications to visualize place connectivity at various geographic levels with downloadable and ready-to-use connectivity matrices to support a wider community research needs. Taking advantage of big social media data and the advancement of high-performance computing, we introduce a place connectivity index (PCI) and an array of PCI datastets based on people's movement among places captured from big Twitter data. Specifically, in this study, we computed global PCI from billions of geotagged tweets aggregated at different geographic levels to reveal place connectivity at multipe scales, including world country (inter-country connectivity), world first-level subdivision (inter-state/province, and intra-country connectivity), US metropolitan area (inter-unban area connectivity), US county (inter-city/county connectivity), and US census tract (intra-city connectivity). We compared population movement derived from Twitter data with the SafeGraph (2020) movement data in the US to evaluate how well geotagged tweets captured population movement. We compared PCI with Facebook's SCI, a popular connectivity index based on social networks, to reveal the association between spatial interactions and social interactions. We also investigated the spatial properties of PCI including distane decay and boundary effect.

The utility of PCI is exemplified in two applications: 1) modeling the spatial spread of COVID-19 during the early stage of the pandemic and 2) modeling hurricane evacuation destination choice. The results demonstrate the great potential of PCI in addressing real-world problems requiring place connectivity knowledge. Finally, we constructed massive PCI matrices and launched an interactive portal for users to visualize the strength of connectivity among geographic regions at various scales. The derived global PCI matrices at various geographic scales are open-sourced to support research needs. Serving as a harmonized and understandable connectivity metric, the multi-scale PCI data with the ability to "zoom in" and "zoom out" are expected to benefit varied domains demanding place connectivity knowledge, such as disease transmission modeling, transportation planning, evacuation simulation, and tourist prediction.

**Place Connectivity Index**

A Place Connectivity Index (PCI) between two places is defined as the normalized number of shared persons (unique Twitter users) between the two places during a specified time period (e.g., one year; Fig. 1). For example, if a user is observed at both places during the time period, the user is considered a shared user between the two places. PCI can be computed at various geographic scales. For example, a place can be a county, state, or country. PCI does not aim to capture the real-time population movement between places (though it is derived from such movement); rather, it provides a relatively stable measurement of how strong two places are connected by spatial interactions. The strength of the connection between two places can be determined by many factors, such as geographic distance (the first law of geography; Miller, 2004), transportation, administrative/regional limits (e.g., states), physical barriers (e.g., rivers and mountains), social networks, demographic and socioeconomic similarities or differences. The shared users among places derived from Twitter data can be considered as an observable outcome of the combined force of these factors, and thus is modeless, with the understanding of Twitter data limitations (e.g., population bias). In this sense, PCI should be calculated in a relatively long time period (e.g., a year) to gather sufficient information to summarize the general patterns.

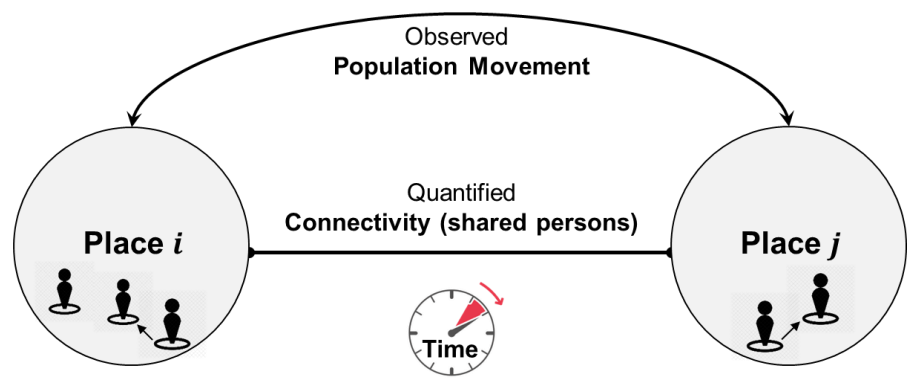

**Fig. 1. Illustration of Place Connectivity Index based on shared social media users.**

Following the general geometric average and normalization strategy (Liu et al., 2014; Bailey et al., 2018; Lin et al., 2019), the PCI between place *i* and place *j* (denoted as *PCI*$_{ij}$) is computed by Eq. 1.

$$PCI_{ij} = \frac{S_{ij}}{\sqrt{S_i S_j}} \quad i,j \in [1,n] \quad \text{Eq. 1}$$

where $S_i$ is the number of observed persons (unique social media users) in place *i* within time period *T*; $S_j$ is the number of observed persons in place *j* within time period *T*; $S_{ij}$ is the number of shared persons between places *i* and *j* within time period *T*; and *n* is the number of places in the study area.

Places with a larger population size tend to have more social media users, and thus tend to have more shared users among them. The denominator in Eq. 1 is used to normalize the metric based on the relative populations in the two places. PCI ranges from 0 to 1. When no shared user is observed between two places, PCI equals 0. If all users in place *i* visit place *j* (vice versa) and the two places have the same number of users (or when *i* = *j* ), PCI equals 1. PCI provides a relative measurement of how strong places are connected through human spatial interactions when assuming all places have the same population (social media users). This allows us to compare PCI among different places to reveal potential spatial, population, and socioeconomic structures. The PCI derived from Eq. 1 is non-directional. The discussion for a directional PCI capturing the asymmetrical connection forces between two places can be found in Appendix A.

# Results

**Global PCI Datasets at Various Geographic Levels**

The computation of PCI is data- and computing-intensive as it involves billions of geotagged tweets and millions of place pairs at various geographic levels. To address this challenge, the computation was performed in a high-performance computing environment (Li et al., 2020a). The steps for computing the 2019 US county level PCI are detailed in Appendix C. With Eq. 1, PCI was computed for the following five geographic levels in this study: 1) worldwide country level for 2019, 2) worldwide first-level subdivision for 2019, 3) US metropolitan area for 2018 and 2019, 4) US county level for 2018 and 2019, and 5) US census tract level for the New York City and Las Angeles County for 2018 and 2019. An interactive web portal was developed to visualize a place's connectivity (PCI) to other places at various geographic levels (Fig. 2, http://gis.cas.sc.edu/GeoAnalytics/pci.html). The following sections report our findings of the PCI properties and potential utility exemplified with the US county level PCI and the world first-level subdivision PCI.

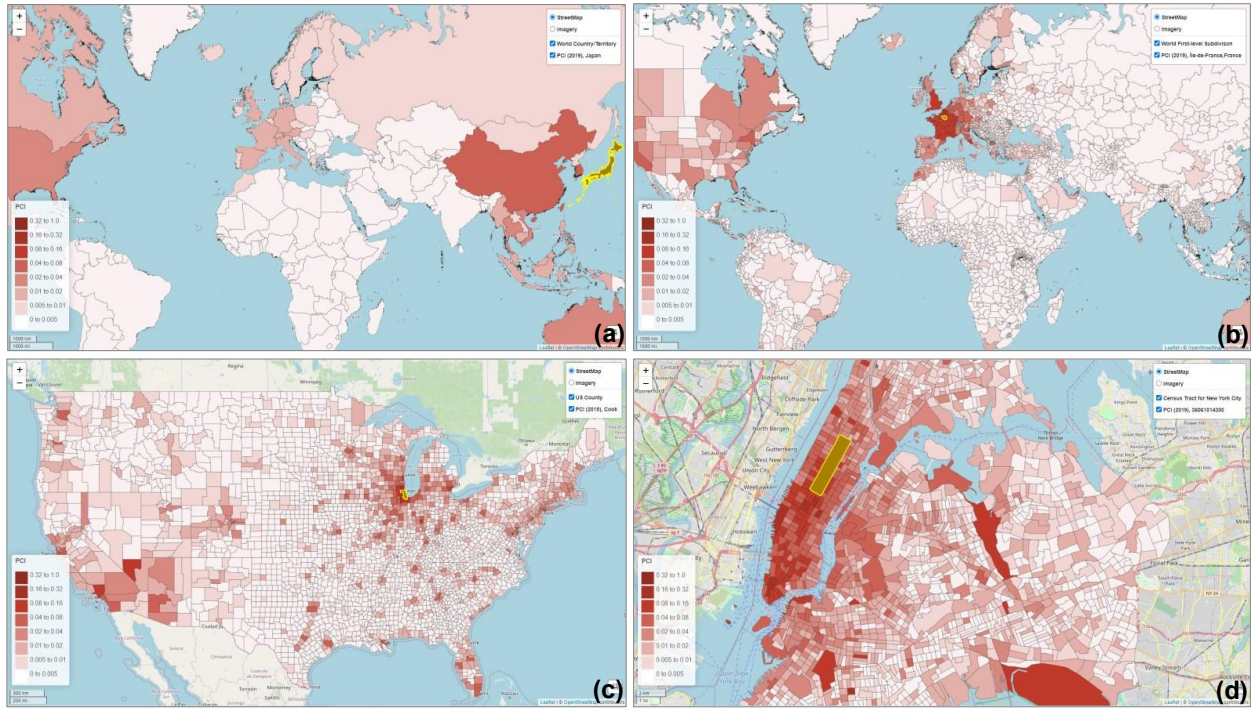

**Fig. 2. Demonstration of PCI at four geographic levels computed with the 2019 global geotagged tweets zoomed in from world country level to US census tract level.** (a) World country level PCI for Japan showing the inter-country connectivity; (b) World first-level subdivision PCI for Ile-de-France (surrounding Paris), France showing the inter-country and intra-country connectivity at the state or province level; (c) US county level PCI for Cook County (Chicago) showing the inter-county/city connectivity; and (d) US census tract level PCI for Central Park, New York City showing the intra-city connectivity.

**Comparing with SafeGraph Population Movement**

One of the key concerns of using social media data (e.g., Twitter) for human mobility studies is its low population penetration rate. For example, only 24% of US adults use Twitter (Pew Research Center, 2019), and the public Twitter API only returns about 1% of the whole Twitter streams. A more detailed descriptive statistics of the collected 2019 worldwide geotagged tweets can be found in Appendix B. Also, Twitter data show bias in its representativeness of population groups. This issue has been examined in a few studies (Jiang et al., 2019; Li et al., 2013; Malik et al., 2015). In light of these issues, it is important to evaluate how well geotagged tweets capture population movements (at the county level in this analysis) since PCI is computed from such movement.

For this purpose, we compared the US county-level population movement derived from Twitter to the movement derived from SafeGraph (https://www.safegraph.com), a commercial data company that aggregates anonymized location data from various sources. According to SafeGraph (2019), the data are aggregated from about 10% of mobile devices (e.g., cellphones) in the US, and the sampling correlates highly with the actual US Census populations, with a Pearson correlation coefficient $r$ of 0.97 at the county level. Specifically, the data we used in this study are the publicly available SafeGraph's Social Distancing Metrics (SDM) (SafeGraph, 2020), a census block group level daily mobility data product going back to January 1, 2019 covering the entire US. Since these data only provide aggregated mobility information, deriving the shared users among counties is not possible. Alternatively, we computed the total number of person-day movements between all contiguous US county pairs in 2019 using the SDM (see Appendix D). To make it comparable, we also computed the total number of person-day movements between all US county pairs in 2019 using Twitter data (see Appendix E). We then compared, using Pearson's $r$, the two person-day movement datasets by county.

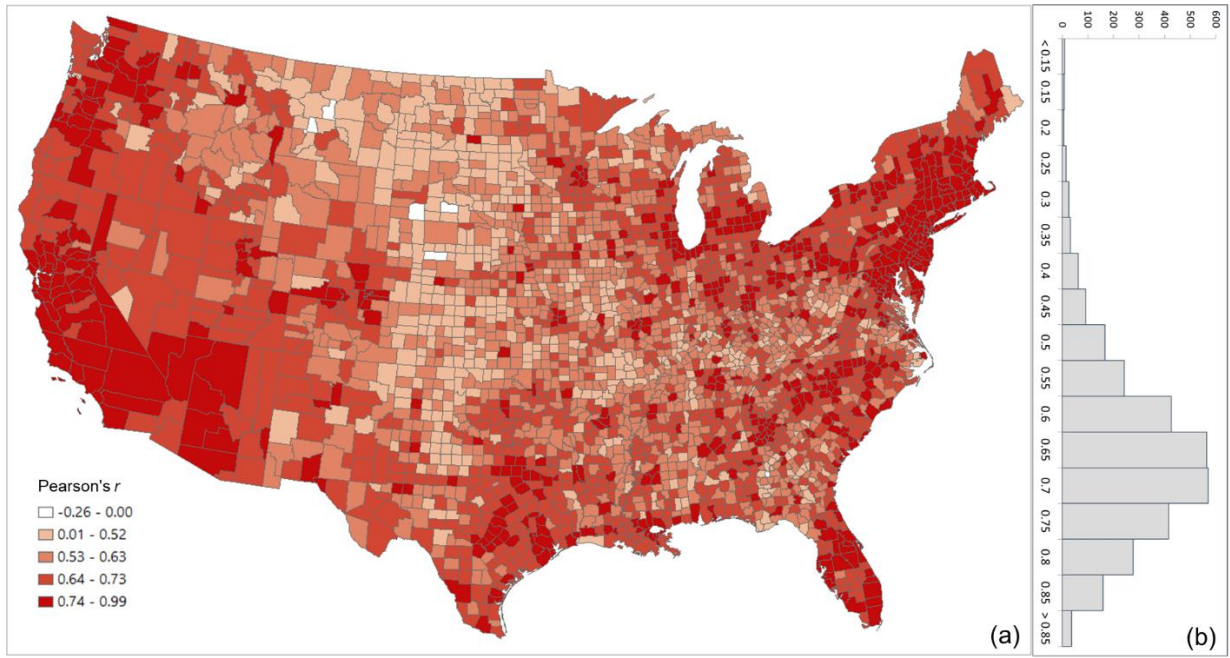

**Fig. 3. Distribution of the Pearson's r between the log Twitter person-day movements and log SafeGraph person-day movements for all counties** (a) Spatial distribution; (b) histogram

The overall Pearson's *r* for all county pairs ($n = 1,516,210$) between log Twitter person-day movements and log SafeGraph person-day movements is 0.71. The rationale for using log transformation (with base 10) is to address the highly skewed distribution of movements among counties. To reveal the spatial variations of the relationship for different areas, we further evaluated the association between the two movement datasets for the county pairs from each county to other counties. The spatial distribution of *r* illustrates lower values generally clustering in less populated areas, such as the Great Plains portion of the US (Fig. 3a). This is as expected, as Twitter data generally suffer in less populated areas due to insufficient tweets collected using the public free API. The histogram (Fig. 3b) indicates the most repeated *r* ranges between 0.65 and 0.75.

To further examine the associations between the two movement datasets and the impact of county population size on the associations, we selected four counties with different geographical contexts and populations ranging from 3,300 to 10,000,000 and plotted the Twitter-derived person day movements and SafeGraph-derived person day movements in 2019 for each county. The scatter plots (Fig. 4) reveal a quasi-linear positive pattern for all four counties. Consistent with Fig. 3, the $r$ value decreases as population decreases for the four counties of Los Angeles County, CA (0.88), Harris County, TX (0.87), Horry County, SC (0.82), and Ford County, KS (0.57). Notably, we observed only a slight drop in $r$ (from 0.88 to 0.82) for Horry County with a relatively small population of 354,081. The findings indicate that geotagged Twitter-derived movement has a strong linear association with SafeGraph-derived population movement and reinforce that geotagged tweets can well capture population movements among places (counties in this analysis).

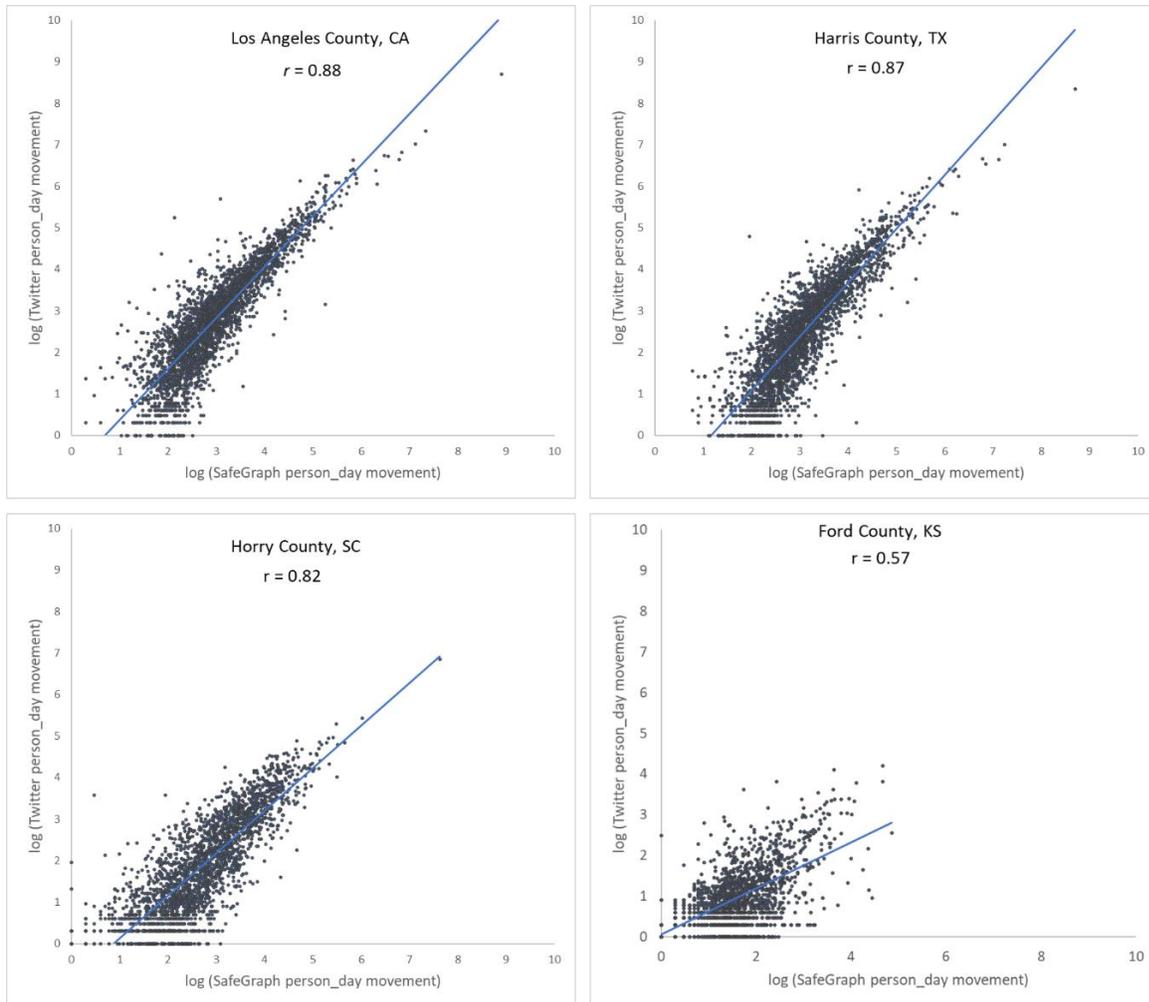

**Fig. 4. Scatter plots of log Twitter-derived person day movements and log SafeGraph-derived person day movements in 2019 for the four selected counties with varying populations.** (a) Los Angeles County, California (CA), including Los Angeles metropolitan area. 2019 population: 10.04 million; (b) Harris County, Texas (TX), including Houston city. The most populous county in TX. 2019 population: 4.71 million; (c) Horry County, South Carolina (SC), including the popular beach destination Myrtle Beach. 2019 population: 354,081; and (d) Ford County, Kansas (KS), including the small Dodge City. 2019 population: 33,619. Population data were derived from the American Community Survey (ACS) 5-Year Data (2015–2019). US Census (2019).

**Comparing PCI with Facebook SCI**

We contrasted the PCI for each of the US counties with the Facebook Social Connectedness Index (SCI) data (Bailey et al., 2018a). This comparison allows us to evaluate the hypothesis that places connected through (social media) friendship links are likely to have more physical interactions (e.g., population movement). This hypothesis has already been suggested in recent studies (Kuchler et al., 2021) but not corroborated using SCI data. Thus, demonstrating this connection is relevant for many reasons, such as understanding spatial behavior under normal circumstances (e.g., business or commercial relationships, tourism, and migrations) or during extraordinary events such as a pandemic (e.g., the spread of infectious diseases) or a natural hazard (e.g., evacuation corridors).

As a measure of social connectedness based on friendship links on Facebook, SCI revealed that the majority of these links are found within 100 miles, showing an intense distance decay effect (Bailey et al., 2018b). The hypothesis of a positive association between social and spatial connections makes intuitive sense and helps understand population dynamics at different scales. To evaluate this, we first analyzed the correlation between PCI and SCI using all county pairs that had both PCI and SCI values ($n = 1,702,531$). Log transformation was used to address the highly skewed distribution of the PCI and SCI values among counties. Note that PCI values were multiplied by 1000 before taking the log to avoid negative values. The overall $r$ of 0.62 indicates a strong linear association between social and spatial connections.

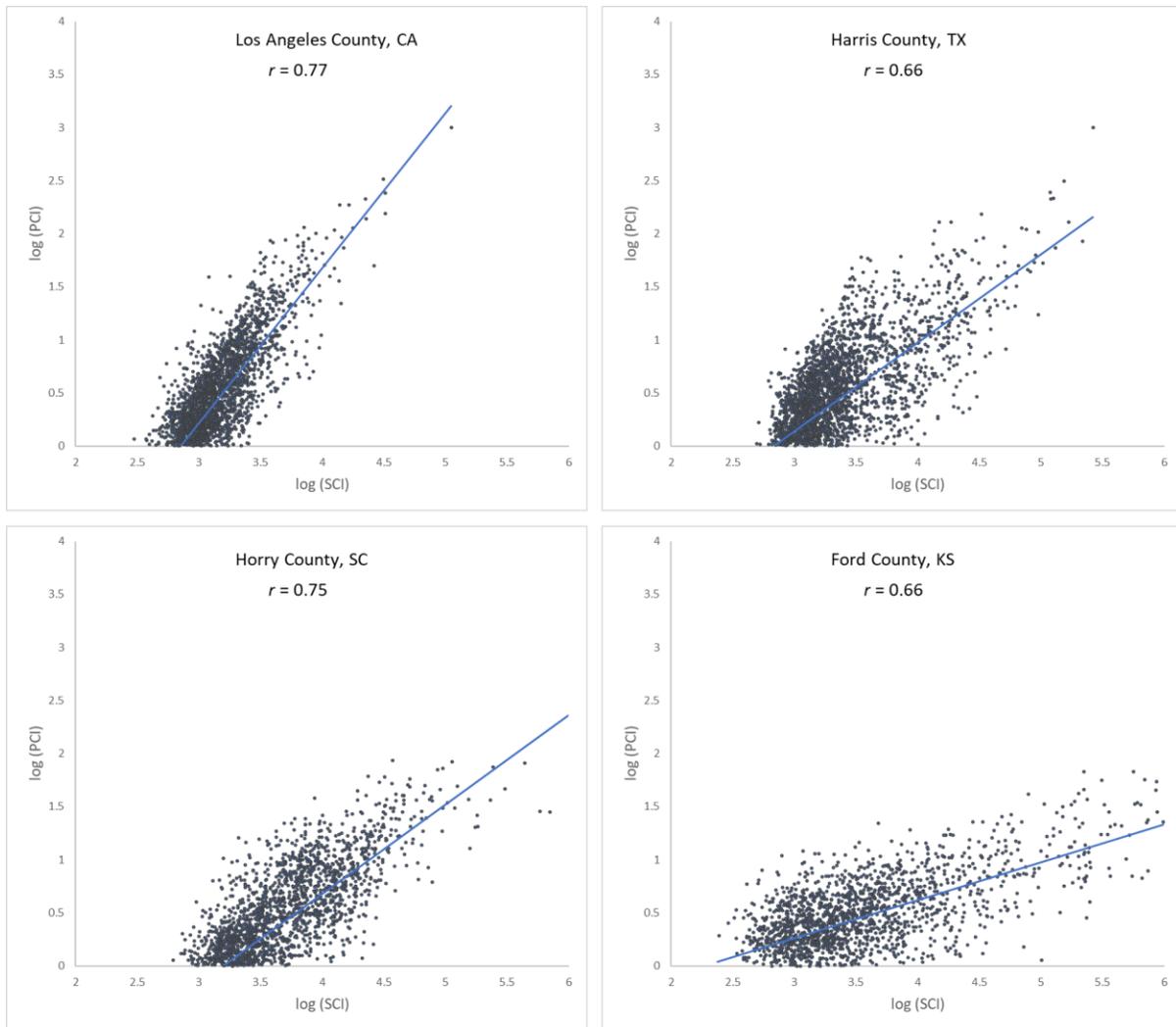

**Fig. 5. Scatter plots of log PCI and log PCI for the four counties**

Fig. 5 shows the scatter plots of log PCI and log SCI in 2019 for the four counties used in the previous section, further confirming the association of a measure of social connectedness with an index of spatial connectivity. The scatter plots also reveal that the association between SCI and PCI is not always stronger in more populated counties (e.g., *r* for Harris County is 0.66 while for the less populated Horry County, it is 0.75). Another interesting observation is that the slope of the best-fit line is higher in more populated counties (e.g., Los Angeles County) than in lowly populated areas (Ford County), indicating that the same amount of change in friendships

(SCI) is associated with a larger change in people's movement (PCI) in more populated counties, vice versa. To further examine the variations of such association among counties, we computed the Pearson's *r* between PCI and SCI for each county to other counties. Fig. 6a shows that strong correlations are generally clustered in Midwest US, Texas, and Southeast Georgia. The histogram (Fig. 6b) shows the most repeated *r* ranging between 0.70 and 0.75.

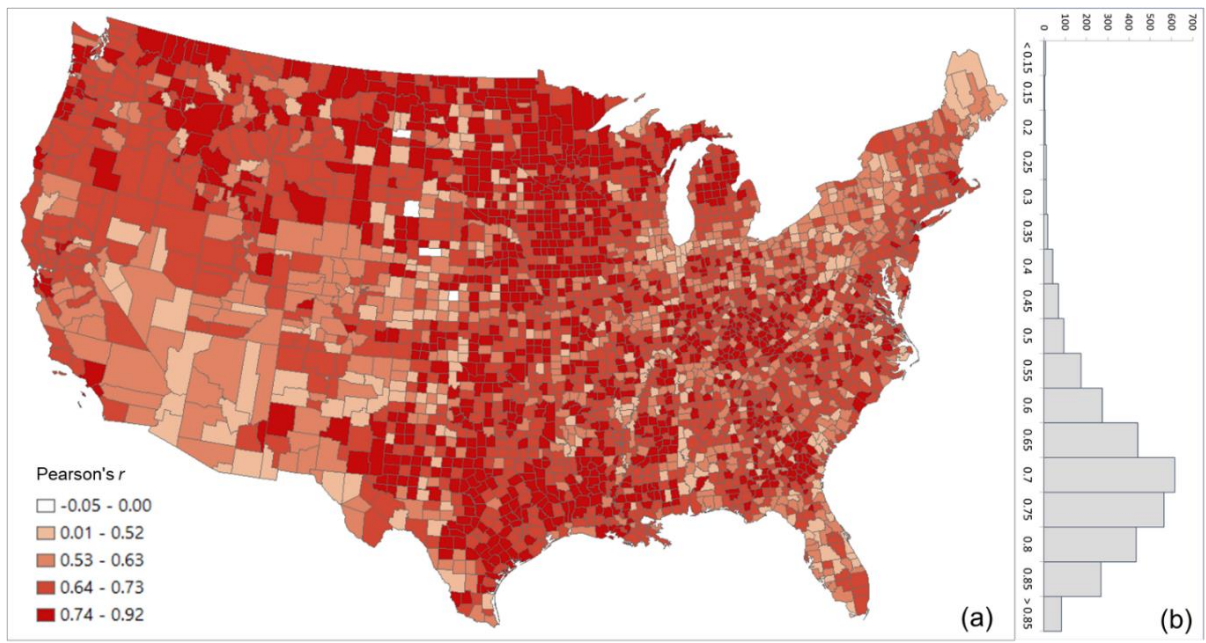

**Fig. 6. Distribution of the Pearson's *r* between log PCI and log Facebook SCI for all counties** (a) Spatial distribution; (b) histogram

The strong association between PCI and SCI confirms the hypothesis that regions connected through (social media) friendship links are likely to have more physical interactions. However, our findings also suggest caution about the relationship between these two variables. Although PCI and SCI are associated, one cannot substitute one for the other, as they represent different phenomena: social versus spatial behavior. More studies are needed to better understand the driving forces (e.g., urban-rural, demographic, and socioeconomic factors) behind the associations of the two variables. We believe PCI is an important addition, as it involves a new standardized measure of spatial connectivity based on population movement.

**Distance Decay Effect**

Our analysis revealed that PCI expresses a clear distance decay effect. In other words, the spatial connectivity between two distant places is likely to be lower than that observed between two near counties. However, there are some nuances in this broad assertion. Fig. 7 illustrates the association between log PCI and the log distance for each county to all other counties. The map (Fig. 7a) shows that less populated (rural) areas of the Midwest, Pacific Northwest, or Texas have a stronger negative association between PCI and distance, meaning that these communities are more tightly knit with surrounding areas than with more distant communities (stronger distance decay effect). This phenomenon is also reflected in Fig. 8, where $R^2$ values of the power-law function decrease dramatically from lowly populated Ford County (0.494) to Harris County (0.154) to highly populated Los Angeles County (0.065). Pearson's *r* was not used as the scatter plot, as the relationship is nonlinear. It should be noted that population size is likely a compounding factor that goes along with urban centers (e.g., metropolis) with large airports.

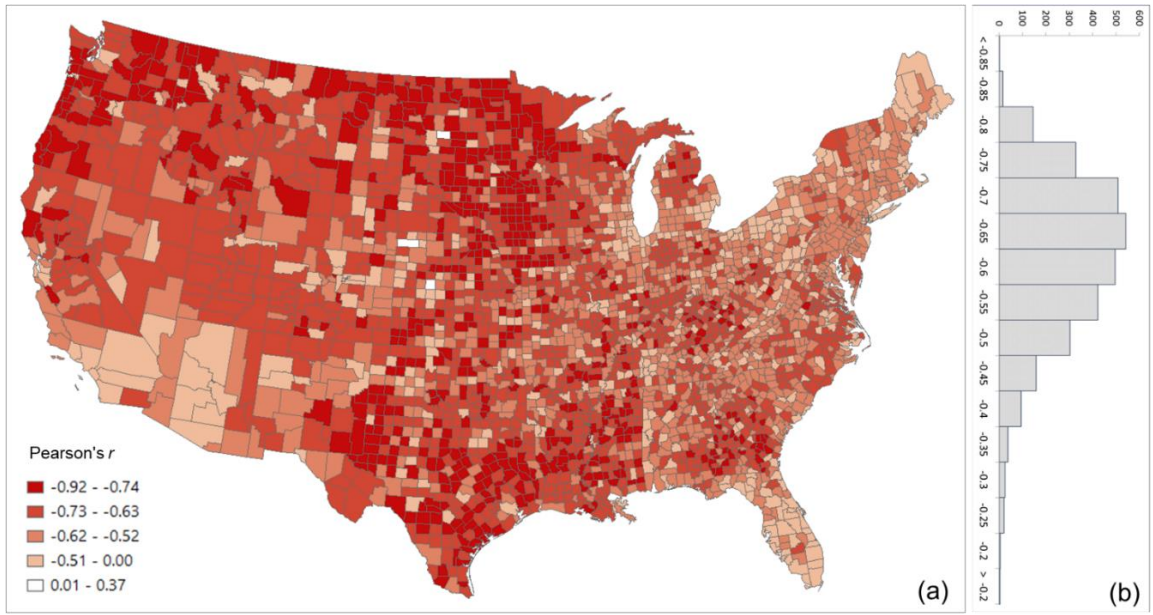

**Fig. 7. Distribution of the Pearson's r between log PCI and log distance for all counties** (a) Spatial distribution, (b) histogram

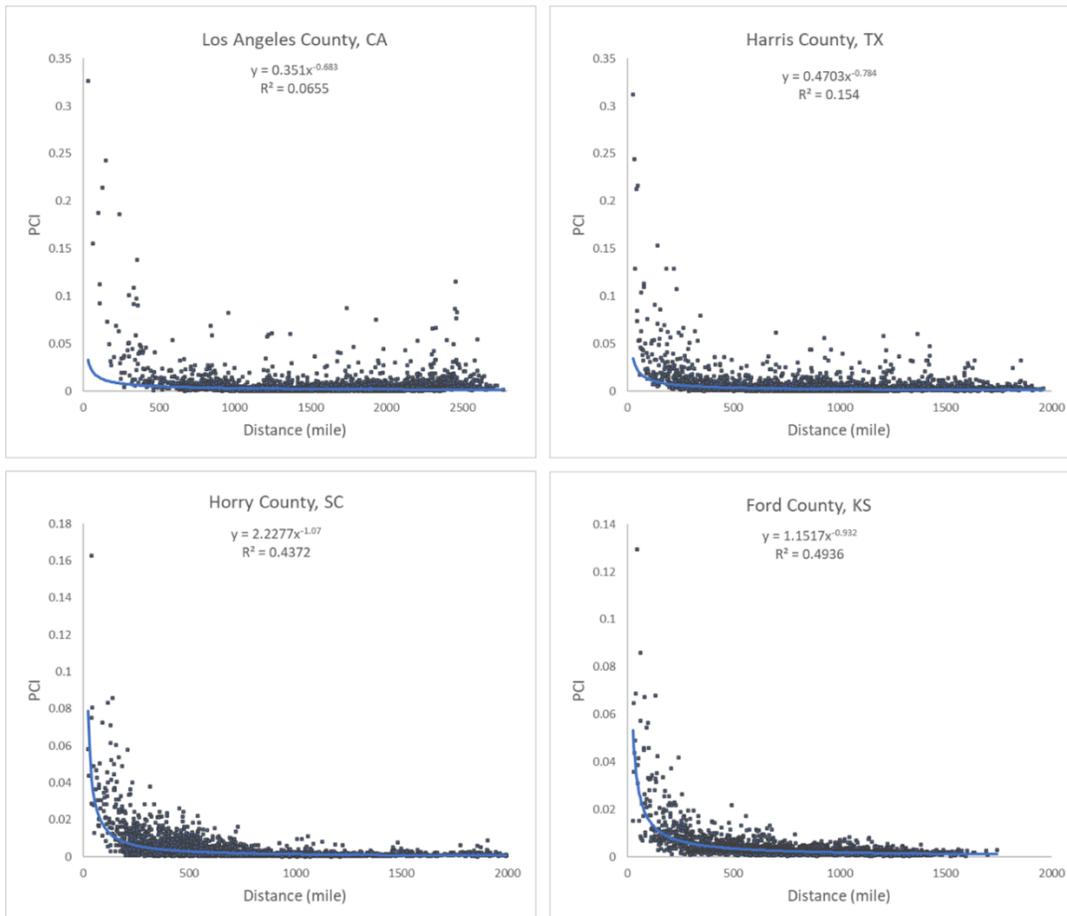

**Fig. 8. Scatter plots of PCI and distance for the four counties.**

The maps in Fig. 9 depict how the selected four counties are connected to other counties based on the PCI, which agrees with the above observations. On the other hand, Fig. 9 also shows that highly populated and touristy urban areas (well connected through airports), such as New York City, Miami, Orlando, Chicago, or Las Vegas, act as poles of attraction for people from distant locations. This is clear in Fig. 9a, where we can see how Los Angeles County, for instance, is more closely linked through spatial interactions with the New York City metropolitan area than with some California or Nevada counties. This behavior is also easily detected in Fig. 8 through the outliers of the point distributions in Los Angeles County.

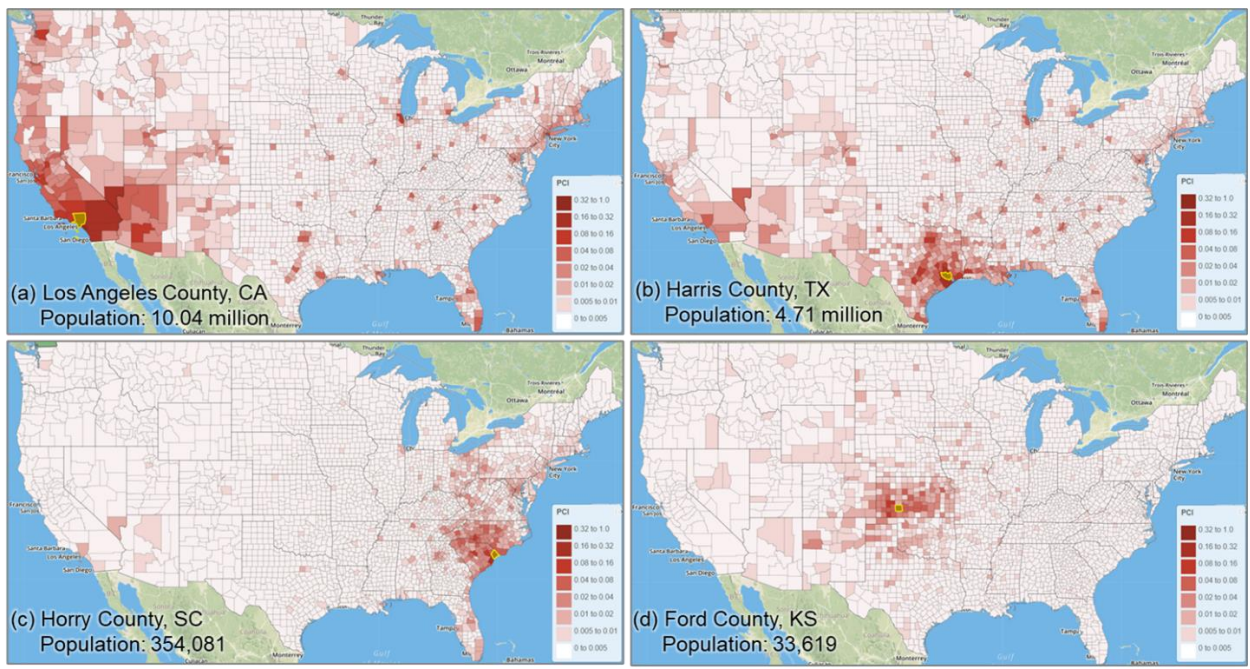

**Fig. 9. The selected four counties (highlighted with yellow boundaries in the maps) and their PCIs with other counties in the contiguous US.** Population data were derived from ACS 5-Year Data (2015–2019) and US Census (2019).

**Boundary Effect**

Inspired by Bailey et al. (2018b), we also considered the effect of administrative borders shaping spatial connectivity. A higher PCI between a county pair indicates a strong relationship geographically. In a general sense, people tend to travel to their adjacent counties more frequently than non-adjacent counties. However, do the residents near the state border prefer the in-state counties as their destinations rather than the adjacent county across the state border? Or are the out-of-state counties more attractive? If state borders have a role in explaining spatial connectivity, people will tend to travel more within their home states than in neighboring states, even when the distance is fixed.

To evaluate the state boundary effect for each of the four counties, we first ran a linear regression with the following variables: the distance between the county and all other counties in the contiguous US (*distance*), a categorical variable (*same_state*), and PCI (as the dependent variable). The result indicates that the *same_state* variable shows a strong positive effect ($p < .001$) on PCI even after controlling for distance (Table 1). This implies that these four counties are more tightly (spatially) connected with other counties within the same state, even when compared to nearby counties in other states.

**Table 1** *Regression Results for the Four Counties using the same_state as an independent variable and PCI as the dependent variable, controlling for the distance between counties.*

|  | Los Angeles County | | Harris County | | Horry County | | Ford County | |
|---|---|---|---|---|---|---|---|---|
|  | Coefficient | SE | Coefficient | SE | Coefficient | SE | Coefficient | S.E. |
| Intercept | 0.0059*** | 0.0008 | -0.0075*** | 0.0007 | 0.0078*** | 0.0003 | -0.0073*** | 0.0003 |
| Same state | 0.0495*** | 0.0017 | 0.0170*** | 0.0010 | 0.0273*** | 0.0010 | 0.0170*** | 0.0006 |
| Distance | -8.3E-07 | -4.5E-07 | -3.6E-06*** | 6.9E-07 | -4.5E-06*** | 2.5E-07 | -4.9E-06*** | 3.7E-07 |
| Adjusted $R^2$ | 0.24 | | 0.16 | | 0.33 | | 0.45 | |
| Observations | 3008 | | 2932 | | 2446 | | 1788 | |

Note: *$p < 0.1$  **$p < 0.05$  ***$p < 0.01$. The county centroid was used for the distance calculation between two counties (distance unit: mile).

To test whether existing state borders are similar to the borders formed when we grouped together the US counties into communities (clusters) based on their spatial connectivity (i.e., PCI), we used a hierarchical agglomerative linkage clustering method following Bailey et al. (2018b) to create such homogenic spatial connectivity communities and compare them with the state administrative division of the US. Hierarchical agglomerative clustering groups county pairs based on their distance in feature space. In our experiment, the "distance" is defined as the inverse of PCI, which means a low PCI in a county pair has a long distance, and vice versa. In the beginning, every county is viewed as a separate community, and the two closest communities are combined into a new community. Distances of combined communities will be updated by the average of distances between county pairs of community pairs. The clustering stops when all counties are combined into a target number of communities. We chose 20, 50, and 75 clusters as the targeted number of communities.

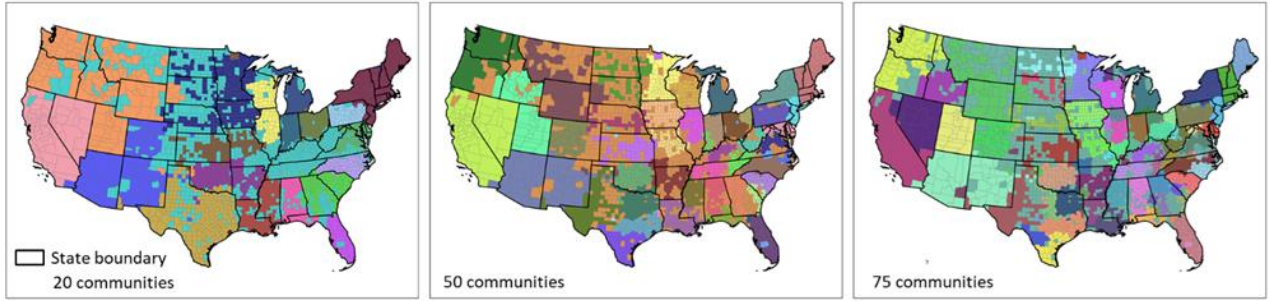

**Fig. 10. Results of the hierarchical agglomerative clustering of PCI with three different targeted numbers of communities for the contiguous US. Each color depicts a community.**

As shown in Fig. 10, most resulting distinct communities in the three maps are spatially contiguous, revealing the strong spatial connectivity of neighboring counties, an obvious consequence of spatial proximity. However, the resemblance of these three maps with state boundaries is quite remarkable across many areas, supporting the assertion that state boundaries do play a decisive role in shaping the spatial behavior of the population. For example, we can see how several clusters in the southwest US are essentially the state boundaries. Also, many other smaller clusters also respect the actual state boundaries. This pattern holds in the three maps with different cluster sizes. A strong state boundary effect was also observed in social connectivity with Facebook SCI (Bailey et al. 2018b). These two findings are likely to be related; however, we do not know which one drives the other or if there are other variables conditioning this behavior (e.g., socio-spatial factors based on institutional or administrative circumstances). Further studies are needed to better understand the state boundary effect of PCI and its connections with SCI.

Fig. 11 shows the hierarchical agglomerative clustering of 2019 PCI for the worldwide first-level subdivisions with two different numbers (25 and 100) of targeted communities (the clustering results for 50 and 200 targeted communities can be found in Appendix F). The country boundaries could be clearly observed in both maps. The results also reveal that the groupings with the 25 communities are consistent with what most people perceive as connected regions (e.g., US with Canada and Europe). However, once into the 100 level, the divisions between east and west start to emerge. Another interesting finding is how unconnected the regions in Africa are, though the country boundary effect is still observable. However, it should be cautious that whether such disconnection resulted from the sparsity of Twitter data in Africa countries (elaborated in the Discussion section) needs further investigation. In summary, the different regions identified in the US and the world using the agglomerative clustering not only demonstrate the boundary effect of PCI, but also suggest that PCI can be potentially used as a tool in regionalization analysis to reveal how places are connected and regions are formed at different geographic levels.

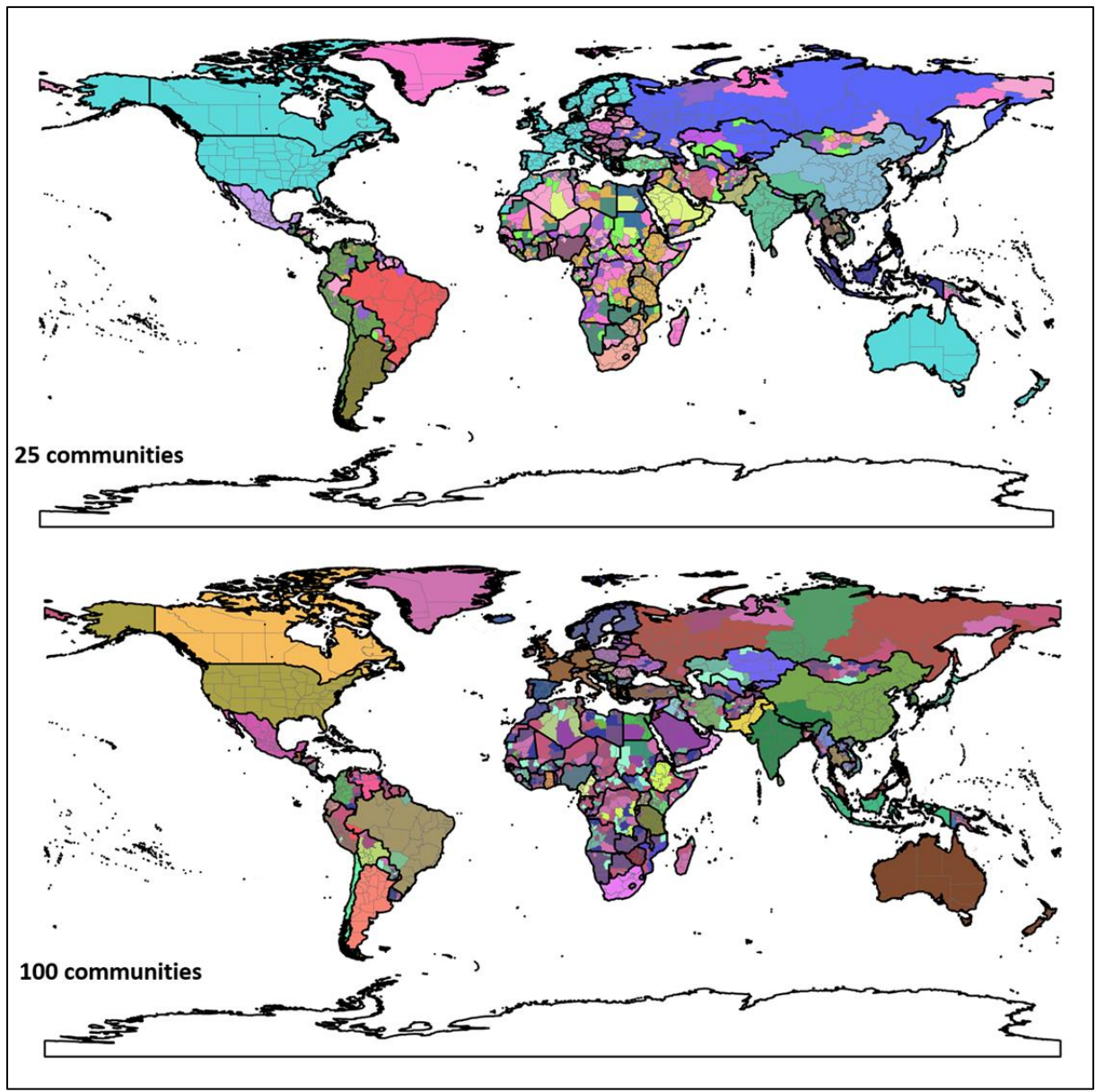

**Fig. 11. Results of the hierarchical agglomerative clustering of PCI with two 25 and 100 targeted numbers of communities for the worldwide country first-level subdivisions.** Each color depicts a community. Boundary data was retrieved from GADM (2018).

**Applications**

PCI can potentially be applied in diverse fields that can benefit from a better understanding of human movement at varying spatial scales, such as infectious disease spread in public health, transportation, tourism, evacuation, and economics. Two examples are provided to exemplify how PCI can be used as a potential factor in analyzing and predicting infectious disease spreading and hurricane evacuation destination choice.

*PCI as a Factor in Predicting the Spatial Spread of COVID-19 During the Early Stage*

Westchester County was an early (March 2020) hotspot of COVID-19 in the US (Hogan et al., 2020). Early confirmed cases and a high infection rate to family and friends increased social tension that residents from Westchester and surrounding areas were reportedly fleeing away (Tully and Stowe, 2020). On the global scale, Lombardy, Italy was an epicenter of the COVID-19, with the first cluster of cases detected on February 21, 2020 (Balmer et al., 2020). The travel restrictions between the US and Europe were not in place until March 12, 2020 (BBC, 2020). In this application example, we explored the relationship between the spread of COVID-19 and PCIs of the two epicenters at the US county level on a regional scale (Westchester County, NY) as well as the state level on a global scale (Lombardy, Italy).

Given that the incubation period of COVID-19 is about two to three weeks (Lauer et al., 2020), the number of cases confirmed before the end of March was used in the later calculation to capture the spread of COVID-19 in early and mid-March for the US county level analysis. Fig. 12 shows the county-level infection rate (number of confirmed cases per 10,000 people) as of March 31, 2020. The number of confirmed cases is based on the New York Times (2020) database, and the total county population is based on the ACS five-year estimation (US Census,

2019). Dark red spots show the hotspots of COVID-19 confirmed cases. Westchester County and surrounding New York City areas were the main hotspots at the end of March.

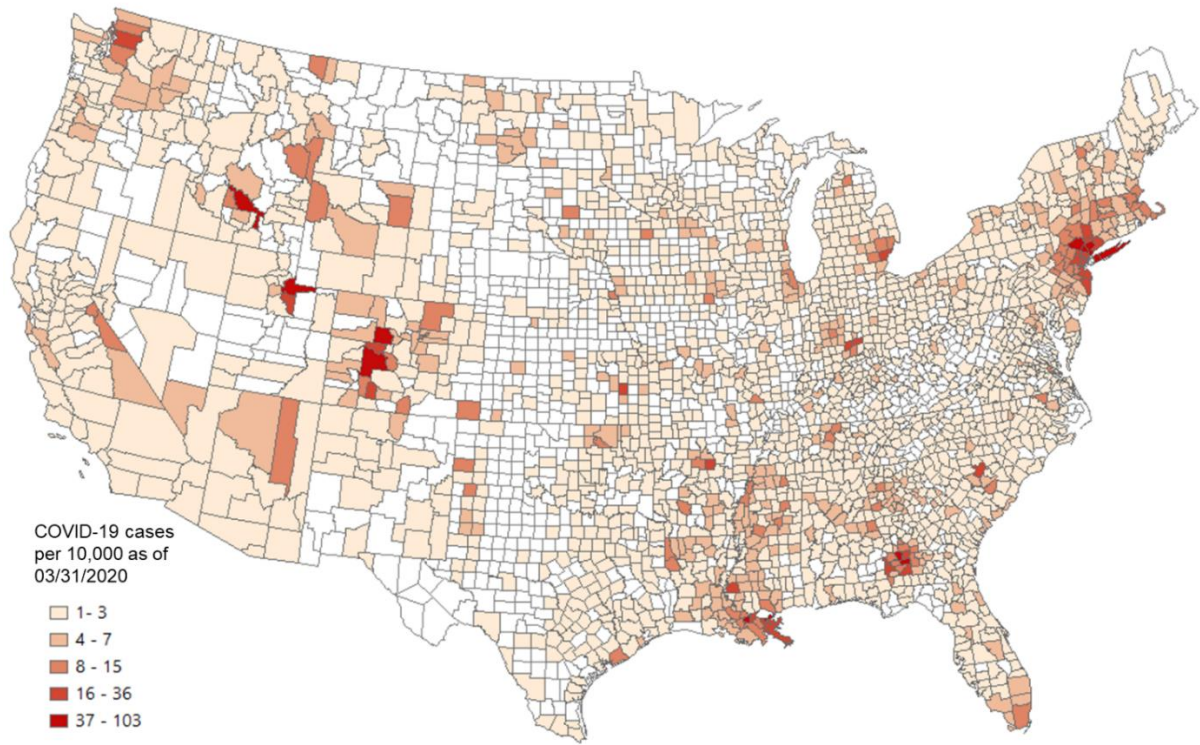

**Fig. 12. US county level COVID-19 cases per 10,000 people as of March 31, 2020.** COVID-19 case data were downloaded from NYT Github (New York Times, 2020). The county population was retrieved from the ACS five-year estimates (2014–2018).

To explore whether outbreaks of COVID-19 in the US are related to people who fled away from New York City in early March (Carey and Glanz, 2020), we used a linear regression model to examine the relationship between COVID-19 infection rate (as a dependent variable) and the connectivity between a given county and Westchester County using four measurements, including PCI computed with 2018 and 2019 Twitter data, respectively, Facebook SCI as of August 2020, and 2020 SafeGraph movement data (the person-day movements computed with the method in Appendix D using data from January to March, 2020). Note that PCI was scaled by 1000 in the regression models to ease the result presentation. Table 2 shows the results of the four linear regression models. For all four measurements, positive relationships are significant at the 0.01 level. Among these four measurements, PCI for both 2018 and 2019 showed the highest adjusted $R^2$ of 0.24 for both years. In other words, 24% of the variance of COVID-19 infection rate in each observed county can be explained by PCI alone. SafeGraph movement results in an adjusted $R^2$ of 0.13. Facebook-based SCI shows the lowest adjusted $R^2$ of 0.08, though the coefficient is still significant ($p < 0.01$).

**Table 2** *Regression Result Using COVID-19 Infection Rate as the Dependent Variable, and PCI, SCI, or SafeGraph as the Predictor Variable.*

|  | 2019 PCI | | 2018 PCI | | Facebook (2020) SCI | | SafeGraph (2020) Person-day movement | |
|---|---|---|---|---|---|---|---|---|
|  | Coefficients | SE | Coefficients | SE | Coefficients | SE | Coefficients | SE |
| Intercept | 1.63496*** | 0.11603 | 1.60351*** | 0.12215 | 2.26827*** | 0.12262 | 2.47287*** | 0.13615 |
| PCI/SCI/SafeGraph | 0.22505*** | 0.00931 | 0.21112*** | 0.00901 | 0.00013*** | 0.00001 | 0.00040*** | 0.00003 |
| Adjusted $R^2$ | 0.24 | | 0.24 | | 0.08 | | 0.13 | |
| Observations | 1847 | | 1755 | | 1847 | | 1497 | |

*$p < 0.1$   **$p < 0.05$   ***$p < 0.01$

Regression models controlling for the effect of geographic distance were also conducted with the four human mobility measurements. Results show that all four measurements still show significant positive correlations with the COVID-19 infection rate ($p < 0.01$; Table 3). The adjusted $R^2$ for SafeGraph-derived movement and Facebook SCI remain unchanged, and the coefficient of the distance variable is not significant ($p > 0.1$). The adjusted $R^2$ for both 2018 and 2019 PCIs only slightly increased by 0.01, from 0.24 to 0.25. While the distance variable is significant in these two models, its impact on the infectious rate is relatively weak given the small coefficient values ($\beta = 0.00087$ for 2019 PCI and $\beta = 0.00091$ for 2018 PCI). We remark that PCI calculated with historical Twitter data of either 2018 or 2019 exhibits similar performance in the two models, suggesting the stability of place connectivity measured by PCI.

**Table 3** *Regression Result Using COVID-19 Infection Rate as the Dependent Variable, and PCI, SCI, or SafeGraph as the Predictor Variable Controlling for Distance*

|  | 2019 PCI | | 2018 PCI | | Facebook (2020) SCI | | SafeGraph (2020) Population movement | |
| --- | --- | --- | --- | --- | --- | --- | --- | --- |
|  | Coefficients | SE | Coefficients | SE | Coefficients | SE | Coefficients | SE |
| Intercept | 0.73883*** | 0.22272 | 0.67638*** | 0.23186 | 2.19947*** | 0.23329 | 2.43267*** | 0.24346 |
| PCI/SCI/SafeGraph | 0.23706*** | 0.00960 | 0.22307*** | 0.00931 | 0.00013*** | 0.00001 | 0.00040*** | 0.00003 |
| Distance | 0.00087*** | 0.00019 | 0.00091*** | 0.00019 | 0.00007 | 0.00020 | 0.00004 | 0.00022 |
| Adjusted $R^2$ | 0.25 | | 0.25 | | 0.08 | | 0.13 | |
| Observations | 1847 | | 1755 | | 1847 | | 1497 | |

*p < 0.1   **p < 0.05   ***p < 0.01

In the global scale analysis, we examined the association between the 2019 US state level PCI with Lombardy, Italy and US state level COVID-19 infection rate (number of cases per 100,000 people) as of March 25, 2020, two weeks after the US placed travel restrictions with Europe. The state level PCI indicates the connectivity strength between each of the 50 US states and Lombardy, Italy (Fig. 13a). As shown in Fig. 13b, PCI with Lombardy exhibited a strong positive association with the US state level COVID-19 infection rate at the early stage of the pandemic ($r = 0.48$, $n = 50$, $p < 0.01$).

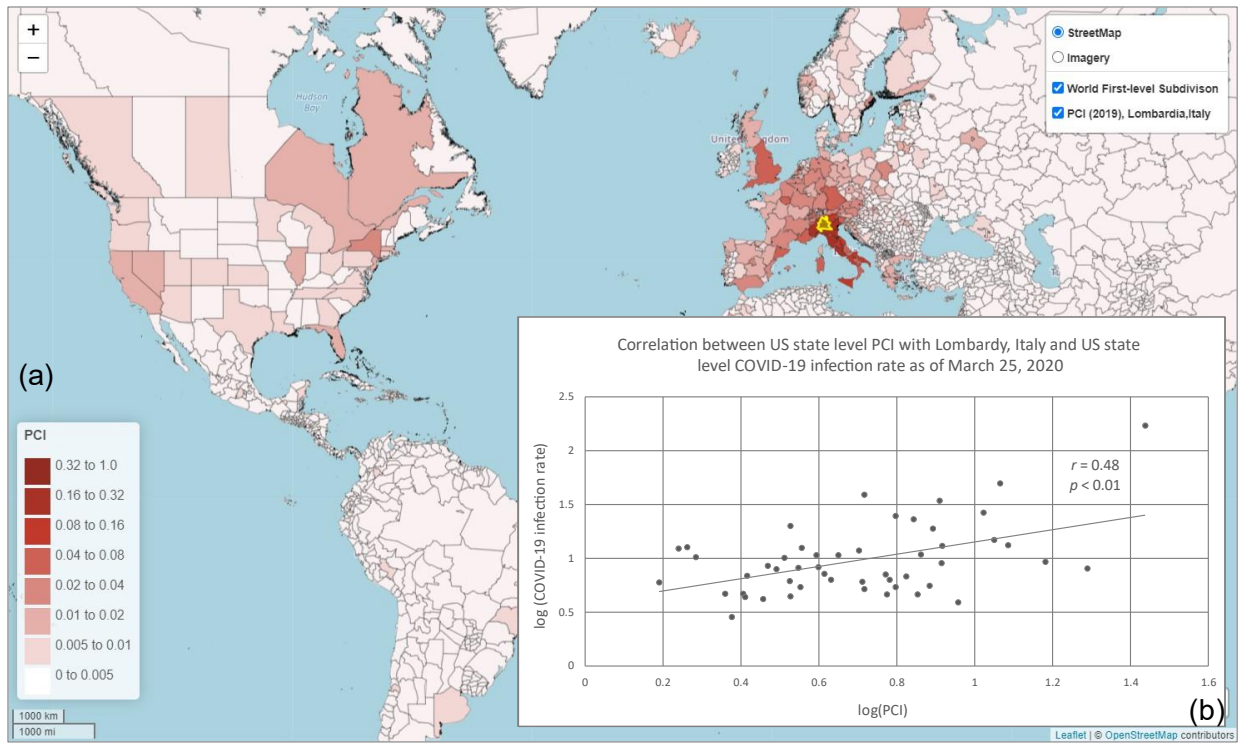

**Fig. 13 Global scale analysis of PCI and COVID-19 infection rate.** (a) Map showing the 2019 world first-level subdivision PCI between Lombardy, Italy and US states (and other parts of the world); (b) Correlation between the log US state level PCI with Lombardy, Italy and log US state level COVID-19 infection rate (number of cases per 100,000 people) as of March 25, 2020. COVID-19 case data were downloaded from NYT Github (New York Times, 2020). The state population was retrieved from the ACS five-year estimates (2014–2018). World first-level subdivision boundary data was retrieved from GADM (2018).

Findings in this application suggest that the multi-scale PCI, computed from historical Twitter data, is a promising indicator in predicting the spatial spread of COVID-19 during the early stage, outperforming more current Facebook SCI (data as of August 2020) and SafeGraph-derived person-day movement data (from January 1 to March 31, 2020) at the US county level.

*PCI as a Factor in Predicting Hurricane Evacuation Destination Choices*

Evacuation of coastal residents has been an effective and important protective action before the arrival of a hurricane (Cutter and Smith, 2009). Understanding where coastal residents are evacuating helps in evacuation route planning and resource allocations (Cheng et al., 2008). Residents of a county are likely to evacuate to a county where they have established relationships (friends, colleagues, familiar lodging stays, etc.). The preexisting relationships would be expressed by the PCI or SCI. In this section, we examined the association between PCI (computed using the 2019 Twitter data) and people's evacuation destination choice using Hurricane Matthew in 2016 as a case study. We hypothesize that people are more likely to evacuate to a county that has a high PCI with the evacuation county (the county being evacuated). For comparison, we also tested the hypothesis that people are more likely to evacuate to a county that has a high SCI with the evacuation county.

Hurricane Matthew was a Category 5 hurricane that visited the east coast of the US at Category 1 in early October 2016. Evacuation orders for coastal counties under potential impact were placed by the governors of Georgia, South Carolina, and North Carolina on October 4, 2016. Twitter users were selected as individual evacuees for testing our hypothesis. The evacuation identification procedure followed the study area and evacuation timeline determined by Martin et al. (2017) and Jiang et al. (2019). In this study, we identified 272 evacuated individual Twitter users from Chatham County, GA, and 241 evacuated users from Charleston County, SC. All selected users had evacuated more than 50 miles away from their original coastal counties, and all of their destinations were not in the potential impact zone. The 272 evacuated individuals leaving Chatham County ended up in 120 destination counties, and the 241 Charleston County evacuees ended up in 118 destination counties (Fig. 14).

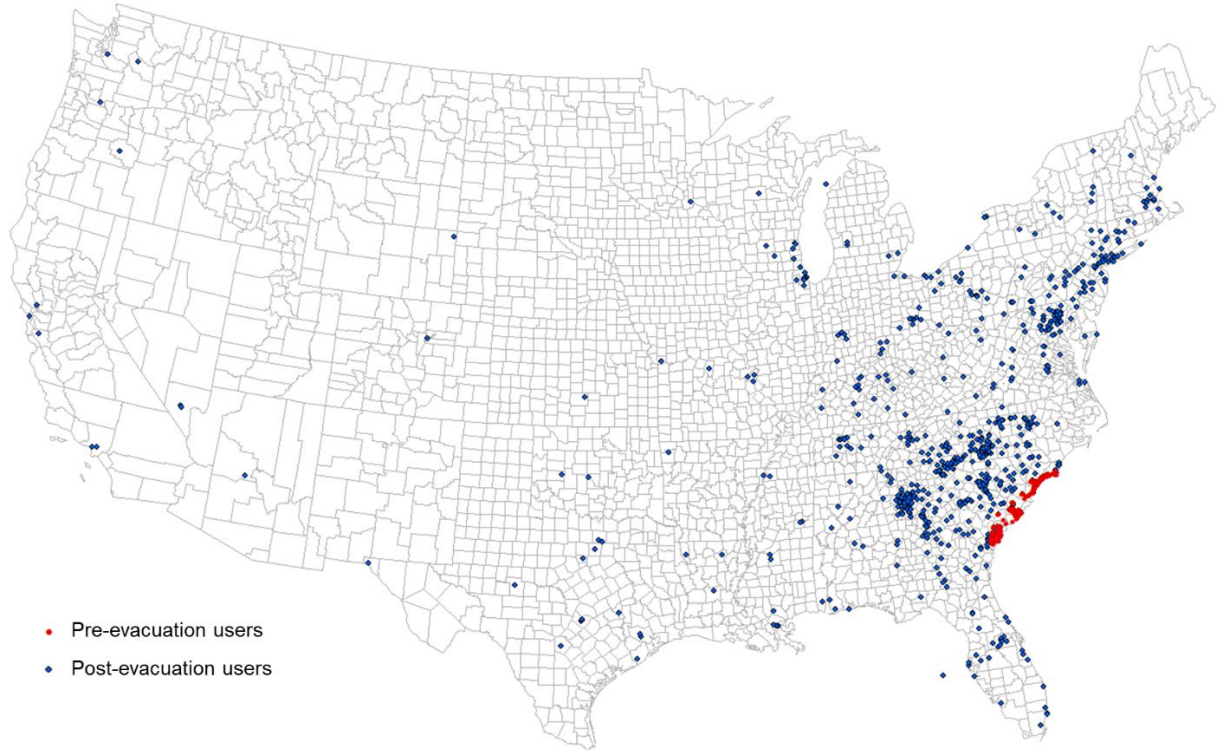

**Fig. 14 Hurricane Matthew Evacuation Estimation Using Geotagged Tweets.** Red dots indicate user locations during the pre-evacuation period (October 2–4, 2016). Blue dots show user locations during the post-evacuation period (October 7–9, 2016).

To test our hypotheses and the potential of PCI in predicting evacuation destination choice, we used linear regression to model the relationship between the number of evacuated users in the destination counties (dependent variable) and PCI of the county pairs between Charleston County (origin) and each of the destination counties ($n = 118$). Note that PCI was scaled by 1000 in the regression models to ease the result presentation. Distance between the evacuation county and each of the destination counties were used in the regression model as controls. SCI was tested by replacing PCI in the regression model for comparison. The same model configuration was used for Chatham County ($n = 120$). Table 4 shows the regression results for the four models.

**Table 4** *Regression Results for the Number of Evacuated Users in the Destination Counties (Dependent Variable) and PCI (or SCI) of the County Pairs Between Evacuation County and Each of the Destination Counties*

|  | Charleston County PCI | | Charleston County SCI | | Chatham County PCI | | Chatham County SCI | |
|---|---|---|---|---|---|---|---|---|
|  | Coefficients | SE | Coefficients | SE | Coefficients | SE | Coefficients | SE |
| Intercept | 0.28676 | 0.23274 | 1.63200*** | 0.32877 | -1.66645** | 0.63781 | 2.57697*** | 0.69610 |
| PCI/SCI | 0.09611*** | 0.00606 | 0.00003*** | 0.00000 | 0.19071*** | 0.01927 | 0.00001 | 0.00001 |
| Distance | 0.00019 | 0.00030 | -0.00047 | 0.00047 | 0.00115 | 0.00083 | -0.00155 | 0.00112 |
| Adjusted $R^2$ | 0.71 | | 0.29 | | 0.47 | | 0.04 | |
| Observations | 118 | | 118 | | 120 | | 120 | |

*$p < 0.1$  **$p < 0.05$  ***$p < 0.01$

For both counties, PCI shows a significant positive association with evacuee counts ($p < 0.01$). SCI shows a significant positive association with evacuee counts for Charleston County ($p < 0.01$), but the coefficient is not significant for Chatham County ($p > 0.1$). The distance variable is not significant for all four models ($p > 0.1$). The PCI model for Charleston County has an adjusted $R^2$ of 0.71, indicating 71% variance can be explained by PCI. However, the adjusted $R^2$ value for SCI has a much lower value of 0.29. For Chatham County, the PCI model has an adjusted $R^2$ of 0.47, while the adjusted $R^2$ value for the SCI model is only 0.04. This application demonstrates the potential of using PCI as a factor in modeling hurricane evacuation destination choice. The comparison of PCI and SCI shows that PCI outperforms SCI in this application scenario.

# Discussions

The evidence of spatial inter-dependency is increasingly apparent across scales, captured by the digital records of growing human mobility and socioeconomic activities. Geotagged social media data records many space-time social contexts where people perceive, act, and interact with each other, allowing researchers to quantify how specific locations are mentioned and related in physical, virtual, and perceived worlds. As a popular social networking platform, Twitter records a substantial portion of human communication and events at various space-time scales. The geotagged tweets can reveal where people visit, with a much larger sample size than conventional surveys. Such large samples can particularly identify the behavior of underrepresented population groups that are not easily accessed by traditional surveys (Hu et al., 2020).

This research employs global geotagged Twitter data to delineate the spatial interactions between places by developing PCI. The results show that geotagged tweets can be used to reveal global place connectivity at various geographic levels. At the US county level, PCI has strong correlations with other data streams such as SafeGraph and Facebook. Compared to the latter two data sources, Twitter data are more openly available overtime at the individual level. The open-sourced global PCI datasets at various geographic levels can thus provide invaluable opportunities to explore human behavior and social phenomena. As demonstrated by the two application examples, PCI can be used for research in infectious disease and hurricane evacuation that benefit from a better understanding of human movement.

The world should be portrayed as networks instead of the mosaic of cities (Beaverstock et al., 2000). As a classical and fundamental research topic, the interactions between locations convey the urban or regional spatial structure. The PCI computed from billions of tweets offers promising opportunities to measure and compare intra- and inter-city connections and flows. Also, PCI can be linked to other large geotagged data, such as Yelp and Transportation Network Company data, to reveal a more completed picture of spatial structure dynamics. Combined with PCI, the place hierarchy and spatial clusters can be revealed based on both virtual and physical interactions.

Although the outcome of the behavior of PCI largely matches our expectations and with the results of other big social data sources, using social media data to identify spatial interaction has the following limitations. We caution that studies using the open-sourced PCI datasets should be aware of such limitations when interpreting the results. First, research using social media has been criticized for being biased for representing specific population groups. For example, young adults are more likely to use Twitter, compared to their older counterparts (Malik et al., 2015; Jiang et al., 2018). Second, the correlation between PCI and other indicators from social networking platforms in the US is largely relevant to the cultural and policy context. Hence, the results may not be readily generalized or used for prediction in other areas. Further studies are needed to evaluate the PCI for geographic areas other than the US. Third, episodic events, such as holidays and hurricanes, would largely attract/hinder users' movement to specific places. It might distort the connectivity if data is only collected for short periods, and thus affect the accuracy and consistency of measurement results. This issue can be addressed by computing PCI over a relatively long period (e.g., one year or longer) or filtering out the data during the affected time period. Lastly, geotagged tweets are unevenly distributed across space and time, which also affects the reliability of such measurements. The data sparsity issue is caused by a variety of factors such as population density, Internet access, and governmental policies on social media.

More studies are needed to evaluate the performance of PCI at different geographic areas and scales by associating and comparing it with other data sources and testing it with other applications.

Despite these limitations, to the best of our knowledge, Twitter data is the most accessible dataset offering the opportunity to extract worldwide human movement at various spatiotemporal scales for a relatively long time period. By open-sourcing the global PCI datasets at various geographic levels, we call for more efforts to tackle these issues and further validate PCI following the suggested future studies and beyond.

## Conclusions

The relationships among places are shaped by dynamic human movement, whose intensity further quantifies the connectivity (strength of the linkages) among places. With the advances in technologies in the past decades, the connectivity among places is ever-evolving dynamically, thus demanding spatiotemporal-continuous observations with harmonized approaches. Fortunately, the emergence of big social media data, benefiting from the advent of geo-positioning techniques and the popularity of social media platforms, offers a new venue where collecting human spatial interactions becomes less-privacy concerning, easily assessable, and harmonized.

In this study, we introduced a global multi-scale place connectivity index based on people's spatial interactions among places revealed from worldwide geotagged Twitter posts. Defined as the normalized number of Twitter users who shared spatial interactions during a specified time period, the proposed PCI is a harmonized and spatiotemporal-continuous place connectivity metric, expected to benefit various domains requiring knowledge in human spatial interactions. The interactive web portal aims to facilitate place connectivity visualization and provide downloadable connectivity matrices to support research needs.

To better understand the characteristics of PCI, we conducted a series of experiments using PCI and other data sources. An overall Pearson's $r$ of 0.71 between the population movement derived from Twitter and SafeGraph (10% penetration in the US population) reveals that geotagged tweets can well capture the population movement at the US county level. The comparison between PCI and Facebook SCI (a popular connectivity index based on social networks) with an overall $r = 0.62$ suggests a strong connection between spatial interactions and social interactions, confirming the hypothesis that "regions connected through many friendship links are likely to have more physical interactions between their residents" (Kuchler, Russel & Stroebel, 2021). Like many connectivity measurements that are bounded by the first law of geography, we found that PCI generally follows distance decay form tested at the county level, while the distance decay effect is found weaker in more urbanized counties with a denser population. This phenomenon can be explained by the existence of long-distance transportation facilitates (e.g., airports, railways, and bus stations) that, to some extent, express a hierarchical diffusion relationship rather than a contagious diffusion. We further observed a strong boundary effect in PCI, indicating that counties in the same state and states/provinces in the same country are more connected, evidenced by their higher PCI values. The different regions identified in the US and the world by using the hierarchical agglomerative clustering suggests that PCI can be used as a tool in regionalization analysis to reveal how places are connected at different geographic levels and scales.

We demonstrated that PCI could address real-world problems requiring place connectivity knowledge using two applications: 1) modeling the spatial spread of COVID-19 during the early stage and 2) modeling hurricane evacuation destination choices. In the first application, we found that the PCI for Westchester County, NY, an early hotspot of COVID-19 in the US, could explain 22% of the variance in COVID-19 cases among US counties at the early outbreak, which was much higher than Facebook SCI (8%) and the population movement derived from SafeGraph (13%). In the global scale analysis, we found that PCI for Lombardy, an early epicenter in Italy, had a strong association with the infection rate at the US state level at the early stage of the pandemic ($r = 0.48$, $n = 50$, $p < 0.01$). In the second application, we found that PCI explains a considerably higher percentage of variance in local residents' choices of destination county during 2016 Hurricane Matthew compared with Facebook's SCI, suggesting the superiority of spatial interactions in modeling evacuation choices than social interactions.

With the effects of geographic distance being weakened by technological advances, place connectivity quantified by human spatial interactions has been evolving since the very first day of modern society and will continue to evolve at an accelerating pace in the future. Taking advantage of the growing popularity of social media, the PCI proposed in this study contributes to a multi-scale, spatiotemporal-continuous measurement of global place connectivity, with the potential to benefit numerous applications such as infectious disease modeling, transportation planning, evacuation modeling, tourism management, to list a few. The methodological and contextual knowledge of PCI, together with the launched visualization platform and open-sourced PCI datasets at various geographic scales, are expected to support research fields in need of prior knowledge in human spatial interactions.

**Data sharing and availability statement:** The following datasets are made available to the public: US census tract level PCI for Los Angeles County and New York City for 2018 and 2019, US county level PCI for 2018 and 2019, world first-level subdivision PCI for 2019, and world country level PCI for 2019. The number of shared users between each place pairs and total number of users for each place are also included in the PCI datasets. The aggregated county-level person-day movements derived from Twitter for 2019 and aggregated county-level person-day movements derived from SafeGraph data for 2019 used in this study are also included. Data download links can be found at https://github.com/GIBDUSC/Place-Connectivity-Index. Facebook SCI data can be downloaded at https://data.humdata.org/dataset/social-connectedness-index. The interactive web portal for visualizing PCI and relevant datasets can be accessed at http://gis.cas.sc.edu/GeoAnalytics/pci.html. Geotagged tweets were retrieved from Twitter using the public free Twitter API (https://developer.twitter.com/en/docs/twitter-api).

**Acknowledgments:** We wanted to thank SafeGraph for open-sourcing their mobility datasets, Facebook SCI authors for open-sourcing their SCI data, and Twitter for providing the free public API for accessing Twitter data streams. The study was supported by the National Science Foundation (NSF) under grant 2028791, the National Institute of Allergy and Infectious Diseases (NIAID) of the National Institutes of Health (NIH) under grant R01AI127203-4S1, and the University of South Carolina COVID-19 Internal Funding Initiative under grant 135400-20-54176. The funders had no role in the study design, data collection and analysis, or preparation of this article.

# Appendices

### A. Directional (or Asymmetrical) PCI

The same number of shared users between two places may have a different impact on each place. Suppose that two places *i* and *j* have 50 shared users, place *i* has 100 users in total, and place j has 1000 users in total. In such a case, place j can be considered to have a larger impact on place i as it has 50% of users shared with place j. Similarly, place i has a smaller impact on place j as it only has 5% of users shared with place i.

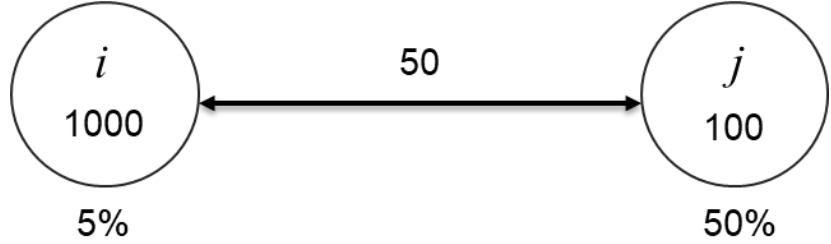

Fig. A1. Illustration of the asymmetrical impact of the shared users on different places

To capture the asymmetrical impact of the same number of shared users on different places, a directional PCI can be derived from Eq. A1 and Eq. A2.

$$PCI_{i \to j} = \frac{S_{ij}}{S_j} \quad i, j \in [1, n] \qquad \text{Eq. A1}$$

$$PCI_{j \to i} = \frac{S_{ij}}{S_i} \quad i, j \in [1, n] \qquad \text{Eq. A2}$$

Where $PCI_{i \to j}$ denotes the impact of place i on place j, and $PCI_{j \to i}$ denotes the impact of place j on place i. $S_i$ is the number of observed persons (unique social media users) in place i within time period T; $S_j$ is the number of observed persons in place j within time period T; $S_{ij}$ is the number of shared persons between places i and j within time period T; and n is the number of places in the study area. For the two places illustrated in Figure A1, the directional $PCI_{i \to j} = 0.500$, $PCI_{j \to i} = 0.050$, and the non-directional $PCI_{ij} = 0.158$.

Same as the PCI (non-directional or symmetrical) discussed in the paper, the directional (or asymmetrical) PCI was also computed for the following four geographic levels: 1) 2019 census tract level PCI for the Las Angeles county, US, 2) 2018 and 2019 county-level PCI for the entire contiguous US, 2) 2019 worldwide first-level subdivision PCI, and 4) 2019 worldwide country level PCI. While the directional PCI makes reasonable sense conceptually, understanding its characteristics and applications requires further investigation. To facility further studies, the derived asymmetrical PCI can be downloaded at https://data.humdata.org/dataset/social-connectedness-index, and be visualized/explored at http://gis.cas.sc.edu/GeoAnalytics/pci.html.

**B. Descriptive Statistics of Worldwide Geotagged Tweets in 2019**

We collected 1,437,611,832 worldwide geotagged tweets in 2019 using the public Twitter Streaming Application Programming Interface (API). Following Martin et al. (2020), we filtered out the non-human tweets (e.g., automated weather reports, job offers, and advertising) by checking the tweet source from which application a tweet is posted. For example, tweets automatically posted for job offers from the source TweetMyJOBS and CareerArc are removed. A list of the Twitter sources that indicate human-posted tweets selected by manual checking is shown in Table B1. Only tweets posted from these sources were included in the analysis and PCI computation. After filtering, a total of 1,409,404,996 geotagged tweets posted by 17,013,612 unique Twitter users were selected. Fig. B1 shows the spatial distribution of these filtered tweets at the country level. The top 40 countries with the most observed unique Twitter users in 2019 are reported in Table B2.

The locations embedded in the geotagged tweets have different spatial resolutions (e.g., exact coordinates, neighborhood, city, and country, etc.). We also analyzed the locations of the

1.4 billion geotagged tweets as the spatial resolution of a geotagged tweet needs to be considered when computing PCIs at different geographic levels. For example, for computing PCI at the US county level, we need to filter out tweets geotagged at a geographic level lower than a city (state level and country level tweets were excluded). As shown in Fig. B2, a majority of tweets (1.1 billion, 79%) were geotagged at the city level, followed by first-level subdivision such as state or province (138.1 million, 9.8%), exact coordinates, (90.4 million, 6.4%), country level (46.2 million, 3.3%), and neighborhood/point of interest (POI) (21.4 million, 1.5%). Based on this, 86.9% of the geotagged tweets can be used for computing the inter-city (county) level PCI. 96.8% of tweets can be used for the first-level subdivision level PCI, and all geotagged tweets can be used for computing country level PCI. For intra-city level PCI (e.g., US census tract level), tweets at the coordinates, neighborhood, or POI levels can be used.

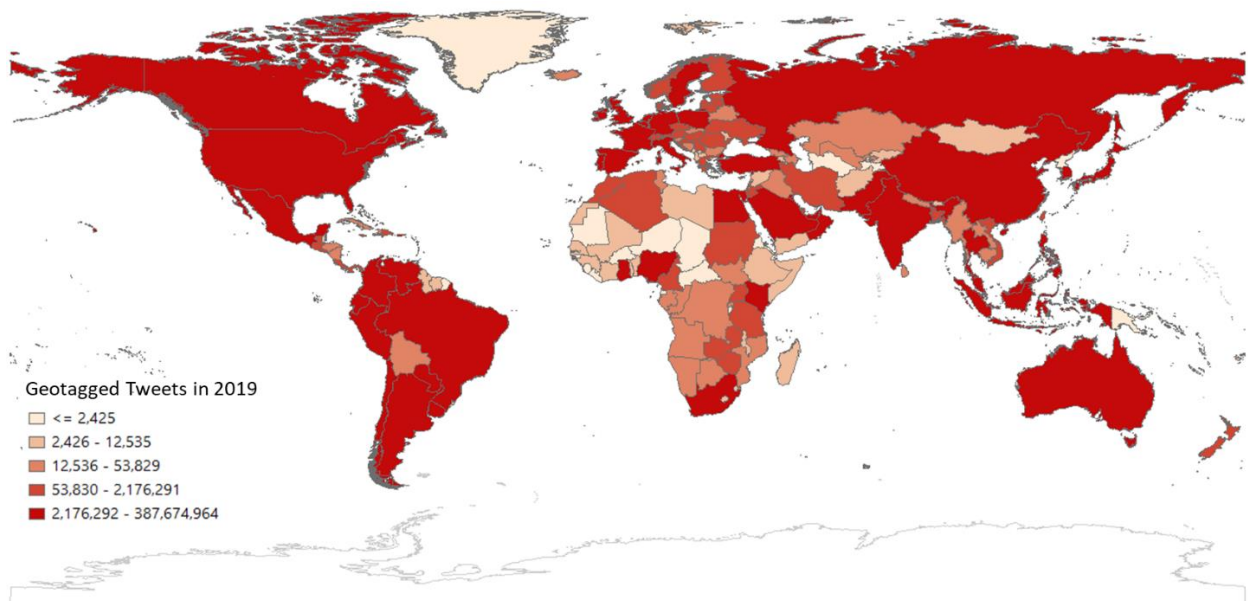

Fig. B1. Spatial distribution of the geotagged tweets at the country level in 2019 (with non-human tweets removed)

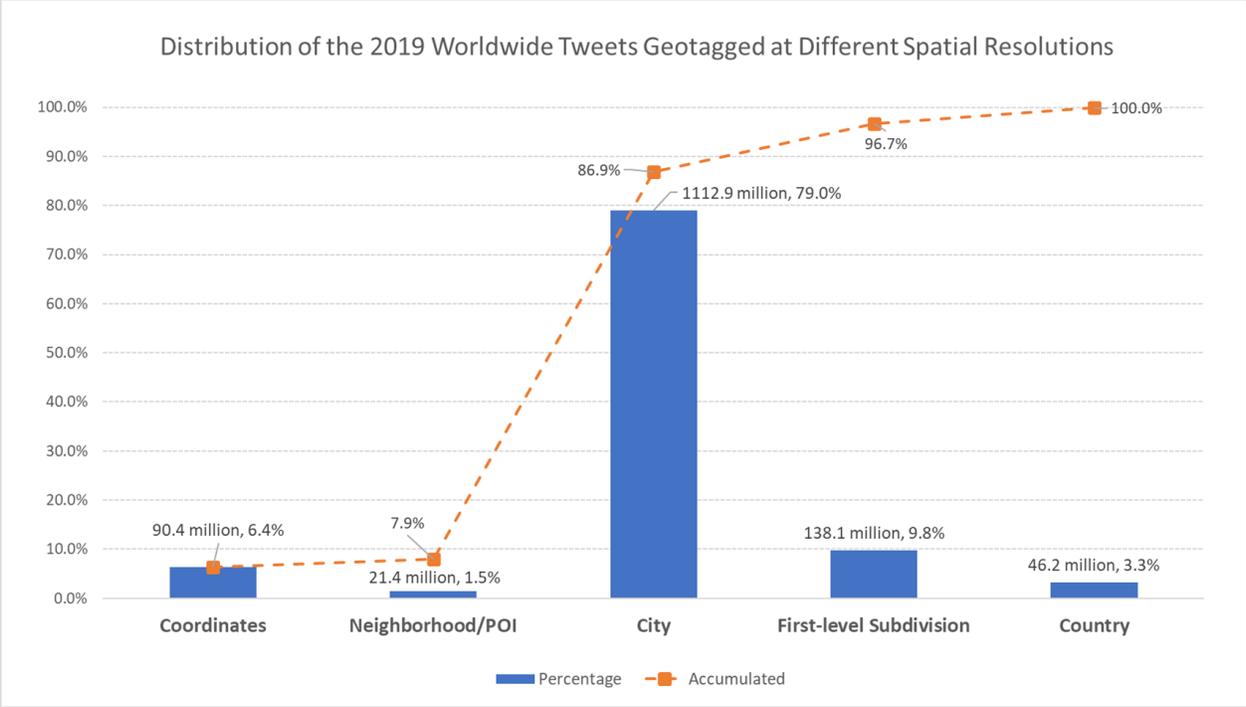

Fig. B2. Distribution of the 2019 worldwide tweets geotagged at different spatial resolutions.

Table B1. List of Twitter sources that indicate human-posted tweets selected by manual checking.

| TweetCaster for Android, TweetCaster for iOS, Tweetings for iPad, Tweetings for Android, Tweetings for Android Holo, Tweetings for Android Tablets, Tweetings for iPhone, Tweetings for iPhone, Tweetlogix, twicca, Twidere for Android #4, Twidere for Android #5, Twidere for Android #7, Twishort Client, Twittelator, Twitter Dashboard for iPhone, Twitter Engage for iPhone, Twitter for Android, Twitter for iPhone, Twitter for Android, Twitter for Android Tablets, Twitter for Apple Watch, Twitter for BlackBerry, Twitter for Calendar, Twitter for iPad, Twitter for iPhone, Twitter for Mac, Twitter for Windows, Twitter for Windows Phone, Untappd, Tweetbot for iOS, Foursquare Swarm, UberSocial for Android, Twitter Web Client, Gay Los Angeles, Gay Santa Monica, Gay West Hollywood, Hootsuite, Instagram, iOS, OS X, |
|---|

> PlumeforAndroid, SoundHound, Squarespace, Talon (Plus), Talon Android, Talon Plus, Echofon, Endomondo, Fenix for Android, Flamingo for Android, Foursquare, Tweet It! for Windows, Tweetbot for iS, Tweetbot for Mac

Table B2. Top 40 countries with most observed unique Twitter users in 2019

| ISO_Code | Users | Tweets |
| --- | --- | --- |
| USA | 4704692 | 387674964 |
| BRA | 1284039 | 215233195 |
| GBR | 1172344 | 70927892 |
| JPN | 1087206 | 84795389 |
| IND | 869366 | 33546715 |
| IDN | 797242 | 40052233 |
| TUR | 717647 | 32230682 |
| ESP | 612306 | 31554779 |
| MEX | 594386 | 29117892 |
| PHL | 571031 | 60788557 |
| SAU | 517797 | 30535954 |
| FRA | 487938 | 21611155 |
| ARG | 410228 | 40673385 |
| CAN | 384644 | 19893446 |
| THA | 349012 | 16894003 |
| ITA | 302492 | 12711959 |
| MYS | 300914 | 23034466 |
| DEU | 273307 | 9885420 |
| COL | 236920 | 14753358 |
| NGA | 218728 | 20002082 |
| NLD | 191777 | 7154017 |
| ZAF | 180112 | 17589164 |
| AUS | 179728 | 9860744 |
| ARE | 172797 | 6222886 |
| EGY | 156265 | 9962841 |
| CHN | 150889 | 2417144 |
| CHL | 143661 | 10082759 |
| RUS | 136230 | 10002772 |
| PRT | 114156 | 6391299 |
| KOR | 111488 | 4700062 |

## C. Computation of US County Level PCI Using 2019 Geotagged Twitter Data

A total of 391,503,203 geotagged tweets from 4,892,458 distinct Twitter users were extracted within the bounding box covering the contiguous United States. For computing PCI at the county level, we filtered out tweets geotagged at a geographic level lower than a city (e.g., state level). If a tweet is geotagged at the place level, the coordinates (latitude and longitude) of the place centroid was used in the analysis. We filtered out the non-human tweets following the procedure introduced in Appendix B. After spatial filtering and non-human filtering, a total of 316,797,441 tweets remained from 4,609,040 unique Twitter users. The process was performed using Apache Impala in a Hadoop environment.

After data cleaning, each Twitter user was assigned to a county based on that user's tweet location daily. For example, if a user tweeted in three counties on a specific day, then for that user on that day, three rows were generated. After processing all users, counties, and days in 2019, a big table (*CountyUserDate*) with three fields (*county, user, date*) was generated, including 95,701,425 county-user-date tuples. The daily table was then aggregated along the date to produce a new table (*CountyUserDays*) with three fields (*county, user, days*), where *days* indicates the number of days (in 2019) a user was observed in a county, including 11,875,433 county-user-days tuples. Based on the *CountyUserDays* table, for each county, two numbers were derived: 1) the number of shared users with other counties and 2) the number of total observed users in each county. Finally, PCI was computed for each county pair (those with shared users) using Eq. 1. This process was conducted using Apache Hive coupled with Esri GIS tools for Hadoop (Esri, 2019).

## D. Computation of the County Person-Day Movements Using 2019 SafeGraph Data

The SafeGraph's Social Distancing Metrics (SDM) data (SafeGraph, 2020) were downloaded and loaded to our Hadoop cluster. There were 23 fields in the SDM table, and three were used to derive the population movement, including *origin_census_block_group*, *destination_cbgs*, and *date_range_start*. The *origin_census_block_group* is the unique 12-digit FIPS code for the Census Block Group. *Destination cbgs* contains a list of key-value pairs with key indicating the destination census block group (from the origin census block group) and "value is the number of devices with a home in *census_block_group* that stopped in the given destination census block group for >1 minute during the time period" (https://docs.safegraph.com/docs/social-distancing-metrics). The *date_range_start* was used to extract the date information.

Based on the three fields, we generated a new table (*SgDailyOD*) with each row showing the number of devices from an original block group to a destination block group on a specific day, resulting in over 6 billion (6,144,802,397) origin-destination flows. Based on the *SgDailyOD* table, we further aggregated the flows to the county level for 2019, generating a new table (*SgCountyPersonDayMovement*) with each row showing the total number of device movements between two counties on all days of 2019, resulting in over 6 million county pairs (6,119,765). Note that the number of movements includes the movements from both directions. For instance, if there are *m* number of movements from county A to county B, and *n* number of movements from county B to county A, then the number of movements between the two counties is calculated as *m* + *n*. The entire process was conducted in our in-house Hadoop cluster using Apache Hive and Impala.

**E. Computation of the County Person-Day Movements Using 2019 Geotagged Twitter Data**

After data cleaning, each Twitter user was assigned to a county based on that user's tweet location daily. For example, if a user tweeted in three counties on a specific day, then for that user on that day, three rows were generated. After processing all users, counties, and days in 2019, a big table (*CountyUserDate*) with three fields (county, user, date) was generated, including 95,701,425 county-user-date tuples. This step was conducted using Apache Hive coupled with ESRI tools for Hadoop. The person-day movement with geotagged tweets was calculated by aggregating the *CountyUserDate* table along the date, generating a new table (*SgCountyPersonDayMovement*) with each row showing the total number of user movements between two counties on all days of 2019, resulting in over 3 million (3,405,113) county pairs.

**F. Results of the hierarchical agglomerative clustering** of 2019 PCI at the worldwide country first-level subdivision level with 50 (Fig. F1) and 200 (Fig. F2) targeted numbers of communities.

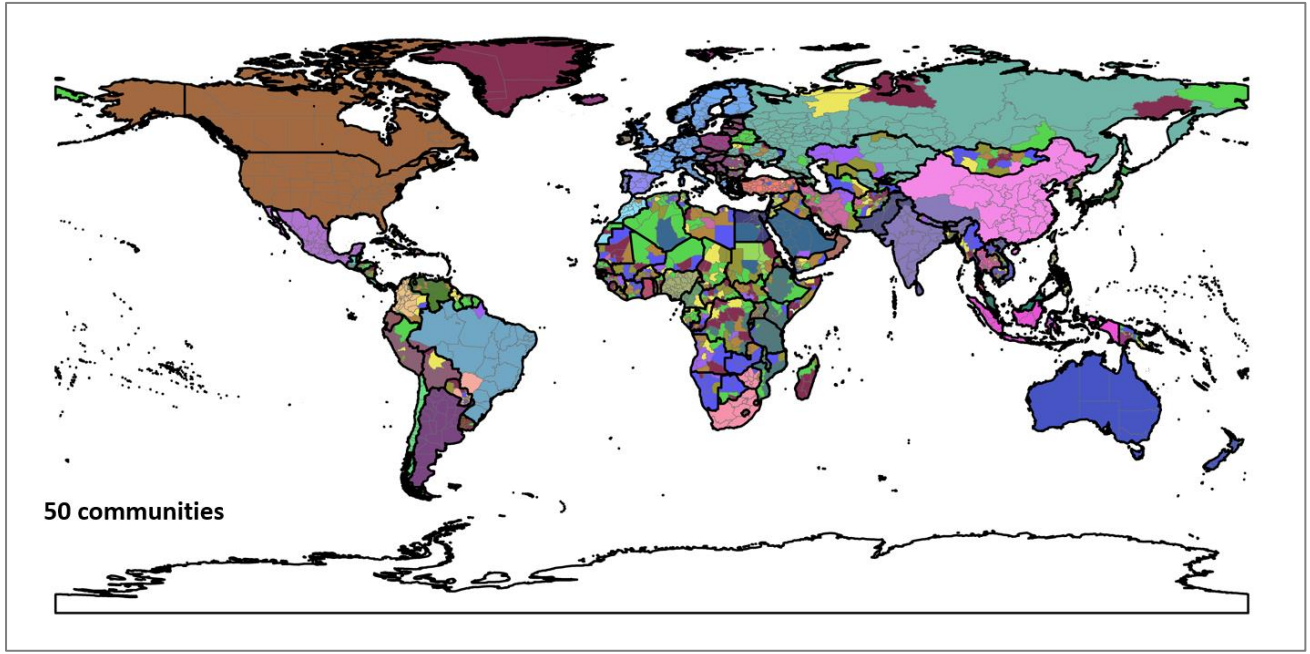

Fig. F1. Clustering result of 50 communities. Each color depicts a community. Boundary data was retrieved from GADM (2018).

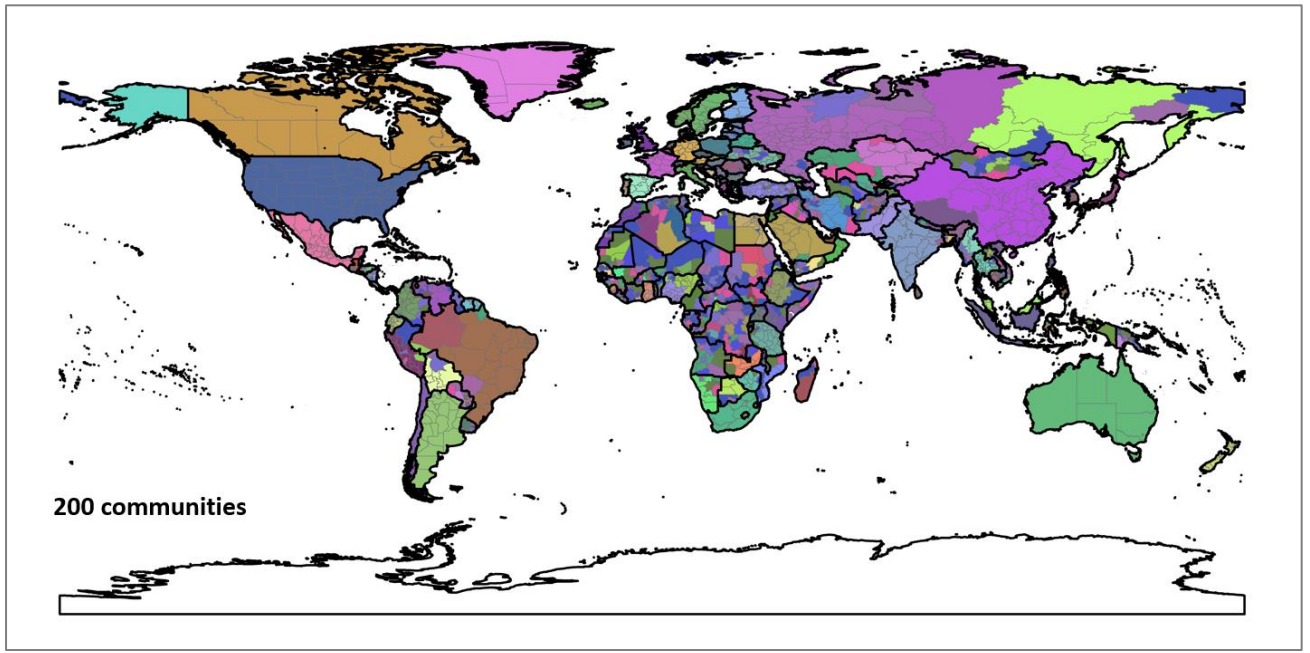

Fig. F2. Clustering result of 200 communities. Each color depicts a community. Boundary data was retrieved from GADM (2018).

# References

Agard, B., Morency, C., & Trépanier, M. (2006). Mining public transport user behaviour from smart card data. IFAC Proceedings Volumes, 39(3), 399–404.

Amini, A., Kung, K., Kang, C., Sobolevsky, S., & Ratti, C. (2014). The impact of social segregation on human mobility in developing and industrialized regions. EPJ Data Science, 3(1), 1–20.

Bailey, M., Cao, R., Kuchler, T., Stroebel, J., & Wong, A. (2018). Social connectedness: Measurement, determinants, and effects. Journal of Economic Perspectives, 32(3), 259–280.

Barcus, H. R., & Brunn, S. D. (2010). Place elasticity: Exploring a new conceptualization of mobility and place attachment in rural America. Geografiska Annaler: Series B, Human Geography, 92(4), 281–295.

Batten, D. F. (2001). Complex landscapes of spatial interaction. The Annals of Regional Science, 35(1), 81–111.

Beaverstock, J. V., Smith, R. G., & Taylor, P. J. (2000). World-city network: A new metageography?

Boyle, P., Halfacree, K., & Robinson, V. (2014). Exploring contemporary migration. Routledge.

Buliung, R. N., & Kanaroglou, P. S. (2006). Urban form and household activity-travel behavior. Growth and Change, 37(2), 172–199.

Balmer, C., Amante, A., Jones, G. (February 21, 2020). "Coronavirus outbreak grows in northern Italy, 16 cases reported in one day". Thomson Reuters. https://www.reuters.com/article/us-china-health-italy-outbreak/coronavirus-outbreak-grows-in-northern-italy-16-cases-reported-in-one-day-idUSKBN20F2GF, last accessed on February 25, 2021

BBC News, Coronavirus: Trump suspends travel from Europe to US, https://www.bbc.com/news/world-us-canada-51846923, last accessed on February 25, 2021

Calantone, R. J., Di Benedetto, C. A., Hakam, A., & Bojanic, D. C. (1989). Multiple multinational tourism positioning using correspondence analysis. Journal of Travel Research, 28(2), 25–32.

Carey, B., & Glanz, J. (2020, May 7). Travel From New York City Seeded Wave of U.S. Outbreaks. The New York Times. https://www.nytimes.com/2020/05/07/us/new-york-city-coronavirus-outbreak.html

Cheng, G., Wilmot, C. G., & Baker, E. J. (2008). A destination choice model for hurricane evacuation. 13–17.

Cutter, S. L., & Smith, M. M. (2009). Fleeing from the hurricane's wrath: Evacuation and the two Americas. Environment: Science and Policy for Sustainable Development, 51(2), 26–36.

de Montjoye, Y.-A., Gambs, S., Blondel, V., Canright, G., De Cordes, N., Deletaille, S., Engø-Monsen, K., Garcia-Herranz, M., Kendall, J., & Kerry, C. (2018). On the privacy-conscientious use of mobile phone data. Scientific Data, 5(1), 1–6.

Derudder, B., & Witlox, F. (2008). Mapping world city networks through airline flows: Context, relevance, and problems. Journal of Transport Geography, 16(5), 305–312.

Durage, S. W., Kattan, L., Wirasinghe, S., & Ruwanpura, J. Y. (2014). Evacuation behaviour of households and drivers during a tornado. Natural Hazards, 71(3), 1495–1517.

ESRI. (2019). GIS Tools for Hadoop: Big data spatial analytics for the Hadoop framework. https://esri.github.io/gis-tools-for-hadoop/

Fiorio, L., Abel, G., Cai, J., Zagheni, E., Weber, I., & Vinué, G. (2017). Using Twitter data to estimate the relationship between short-term mobility and long-term migration. 103–110.

Gao, S. (2015). Spatio-temporal analytics for exploring human mobility patterns and urban dynamics in the mobile age. Spatial Cognition & Computation, 15(2), 86–114.

Ghafoor, H., Koo, I., & Gohar, N.-D. (2014). Neighboring and connectivity-aware routing in VANETs. The Scientific World Journal, 2014.

Gonzalez, M. C., Hidalgo, C. A., & Barabasi, A.-L. (2008). Understanding individual human mobility patterns. Nature, 453(7196), 779–782.

Goodchild, M. F. (2007). Citizens as sensors: The world of volunteered geography. GeoJournal, 69(4), 211–221.

Hawelka, B., Sitko, I., Beinat, E., Sobolevsky, S., Kazakopoulos, P., & Ratti, C. (2014). Geo-located Twitter as proxy for global mobility patterns. Cartography and Geographic Information Science, 41(3), 260–271.

Hu, L., Li, Z., & Ye, X. (2020). Delineating and modeling activity space using geotagged social media data. Cartography and Geographic Information Science, 47(3), 277–288.

Hu, Y., Ye, X., & Shaw, S. L. (2017). Extracting and analyzing semantic relatedness between cities using news articles. International Journal of Geographical Information Science, 31(12), 2427-2451.

Huang, X., Li, Z., Jiang, Y., Li, X., & Porter, D. (2020). Twitter reveals human mobility dynamics during the COVID-19 pandemic. PloS One, 15(11), e0241957.

Huang, X., Li, Z., Jiang, Y., Ye, X., Deng, C., Zhang, J., & Li, X. (2020). The characteristics of multi-source mobility datasets and how they reveal the luxury nature of social distancing in the US during the COVID-19 pandemic. MedRxiv.

Jiang, Y., Li, Z., & Cutter, S. L. (2019). Social network, activity space, sentiment, and evacuation: What can social media tell us? Annals of the American Association of Geographers, 109(6), 1795–1810.

Jiang, Y., Li, Z., & Ye, X. (2019). Understanding demographic and socioeconomic biases of geotagged Twitter users at the county level. Cartography and Geographic Information Science, 46(3), 228–242.

Jurdak, R., Zhao, K., Liu, J., AbouJaoude, M., Cameron, M., & Newth, D. (2015). Understanding human mobility from Twitter. PloS One, 10(7), e0131469.

Kontokosta, C. E., & Johnson, N. (2017). Urban phenology: Toward a real-time census of the city using Wi-Fi data. Computers, Environment and Urban Systems, 64, 144–153.

Kuchler, T., Russel, D., & Stroebel, J. (2021). The geographic spread of COVID-19 correlates with the structure of social networks as measured by Facebook. Journal of Urban Economics, 103314.

Lauer, S. A., Grantz, K. H., Bi, Q., Jones, F. K., Zheng, Q., Meredith, H. R., Azman, A. S., Reich, N. G., & Lessler, J. (2020). The incubation period of coronavirus disease 2019 (COVID-19) from publicly reported confirmed cases: Estimation and application. Annals of Internal Medicine, 172(9), 577–582.

Li, L., Goodchild, M. F., & Xu, B. (2013). Spatial, temporal, and socioeconomic patterns in the use of Twitter and Flickr. Cartography and Geographic Information Science, 40(2), 61–77.

Li, Z., Huang, Q., & Emrich, C. T. (2019). Introduction to social sensing and big data computing for disaster management.

Li, Z., Li, X., Porter, D., Zhang, J., Jiang, Y., Olatosi, B., & Weissman, S. (2020a). Monitoring the Spatial Spread of COVID-19 and Effectiveness of Control Measures Through Human Movement Data: Proposal for a Predictive Model Using Big Data Analytics. JMIR Research Protocols, 9(12), e24432.

Lin, J., Wu, Z., & Li, X. (2019). Measuring inter-city connectivity in an urban agglomeration based on multi-source data. *International Journal of Geographical Information Science*, *33*(5), 1062-1081.

Liu, Y., Wang, F., Kang, C., Gao, Y., & Lu, Y. (2014). Analyzing Relatedness by Toponym Co-Occurrences on Web Pages. *Transactions in GIS*, *18*(1), 89-107.

Liu, Y., Liu, X., Gao, S., Gong, L., Kang, C., Zhi, Y., ... & Shi, L. (2015). Social sensing: A new approach to understanding our socioeconomic environments. Annals of the Association of American Geographers, 105(3), 512-530.

Li, Z., Huang, X., Ye, X., & Li, X. (2020b). ODT flow explorer: Extract, query, and visualize human mobility. arXiv preprint arXiv:2011.12958.

Massey, D. (1994). A global sense of place, in Space, Place and Gender. Cambridge: Polity Press, pp. 146–156.

Ma, X., Wu, Y.-J., Wang, Y., Chen, F., & Liu, J. (2013). Mining smart card data for transit riders' travel patterns. Transportation Research Part C: Emerging Technologies, 36, 1–12.

Macdonald, K., & Grieco, M. (2007). Accessibility, mobility and connectivity: The changing frontiers of everyday routine. Mobilities, 2(1), 1–14.

Malik, M. M., Lamba, H., Nakos, C., & Pfeffer, J. (2015). Population bias in geotagged tweets. People, 1(3,759.710), 3–759.

Martín, Y., Cutter, S. L., & Li, Z. (2020). Bridging twitter and survey data for evacuation assessment of Hurricane Matthew and Hurricane Irma. Natural Hazards Review, 21(2), 04020003.

Martin, Y., Cutter, S. L., Li, Z., Emrich, C. T., & Mitchell, J. T. (2020). Using geotagged tweets to track population movements to and from Puerto Rico after Hurricane Maria. Population and Environment, 42(1), 4–27.

Martín, Y., Li, Z., & Cutter, S. L. (2017). Leveraging Twitter to gauge evacuation compliance: Spatiotemporal analysis of Hurricane Matthew. PLoS One, 12(7), e0181701.

Massey, D. (2010). A global sense of place. Aughty. org.

New York Times. (2021). Coronavirus (Covid-19) Data in the United States. https://github.com/nytimes/covid-19-data

O'reilly, T. (2007). What is Web 2.0: Design patterns and business models for the next generation of software. Communications & Strategies, 1, 17.

Pereira, F., Carrion, C., Zhao, F., Cottrill, C. D., Zegras, C., & Ben-Akiva, M. (2013). The future mobility survey: Overview and preliminary evaluation. 9, 1–13.

Perrin, A., & Anderson, M. (2019, April 10). Share of U.S. adults using social media, including Facebook, is mostly unchanged since 2018. Pew Research Center. https://www.pewresearch.org/fact-tank/2019/04/10/share-of-u-s-adults-using-social-media-including-facebook-is-mostly-unchanged-since-2018/

Perttunen, M., Kostakos, V., Riekki, J., & Ojala, T. (2014). Spatio-temporal patterns link your digital identities. Computers, Environment and Urban Systems, 47, 58–67.

Pham, E. O., Emrich, C. T., Li, Z., Mitchem, J., & Cutter, S. L. (2020). Evacuation departure timing during Hurricane Matthew. Weather, Climate, and Society, 12(2), 235–248.

SafeGraph. (2020). Social Distancing Metrics. SafeGraph. https://docs.safegraph.com/docs/social-distancing-metrics

Salt, J. (1987). Contemporary trends in international migration study. International Migration (Geneva, Switzerland), 25(3), 241–251.

Santos, A., McGuckin, N., Nakamoto, H. Y., Gray, D., & Liss, S. (2011). Summary of travel trends: 2009 national household travel survey. United States. Federal Highway Administration.

Siebeneck, L. K., & Cova, T. J. (2012). Spatial and temporal variation in evacuee risk perception throughout the evacuation and return-entry process. Risk Analysis: An International Journal, 32(9), 1468–1480.

Soliman, A., Soltani, K., Yin, J., Padmanabhan, A., & Wang, S. (2017). Social sensing of urban land use based on analysis of Twitter users' mobility patterns. PloS One, 12(7), e0181657.

Squire, R. F. (2019, October 17). What about bias in the SafeGraph dataset? https://www.safegraph.com/blog/what-about-bias-in-the-safegraph-dataset

Stewart, J. Q. (1948). Demographic gravitation: Evidence and applications. Sociometry, 11(1/2), 31–58.

Tranos, E., & Nijkamp, P. (2013). The death of distance revisited: Cyber-place, physical and relational proximities. Journal of Regional Science, 53(5), 855–873.

Tully, T., & Stowe, S. (2020, March 25). The Wealthy Flee Coronavirus. Vacation Towns Respond: Stay Away. The New York Times. https://www.nytimes.com/2020/03/25/nyregion/coronavirus-leaving-nyc-vacation-homes.html

U.S. Census Bureau. (2019, December 10). American Community Survey 5-Year Data (2009-2019). The United States Census Bureau. https://www.census.gov/data/developers/data-sets/acs-5year.html

Xu, Z., & Harriss, R. (2008). Exploring the structure of the US intercity passenger air transportation network: A weighted complex network approach. GeoJournal, 73(2), 87.

Yang, Y., Li, D., & Li, X. (2019). Public transport connectivity and intercity tourist flows. Journal of Travel Research, 58(1), 25–41.

Ye, X., Gong, J., & Li, S. (2020). Analyzing Asymmetric City Connectivity by Toponym on Social Media in China. Chinese Geographical Science, 1-13.

Zhang, X., Chen, S., Luan, X., & Yuan, M. (2020). Understanding China's city-regionalization: Spatial structure and relationships between functional and institutional spaces in the Pearl River Delta. Urban Geography, 1–28.

GADM, (2018), GADM data at https://gadm.org/data.html, last accessed on January 5, 2021